\documentclass[pdflatex,sn-mathphys-num]{sn-jnl}

\usepackage{lineno}
\usepackage{graphicx}
\usepackage{multirow}
\usepackage{amsmath,amssymb,amsfonts}%
\usepackage{amsthm}
\usepackage{mathrsfs}
\usepackage[title]{appendix}
\usepackage{xcolor}
\usepackage{textcomp}
\usepackage{manyfoot}
\usepackage{booktabs}
\usepackage{algorithm}
\usepackage{algorithmicx}
\usepackage{algpseudocode}
\usepackage{listings}
\usepackage{booktabs}
\usepackage{makecell}
\usepackage{xcolor}
\usepackage{graphicx}

\definecolor{darkgreen}{RGB}{0,100,0}

\newcommand{\gain}[1]{\textcolor{darkgreen}{$\uparrow #1$}}
\newcommand{\loss}[1]{\textcolor{red}{$\downarrow #1$}}
\newcommand{\nochange}{\textcolor{black}{$0\%$}}
\newcommand{\egain}{\textcolor{darkgreen}{$\uparrow$}}
\newcommand{\eloss}{\textcolor{red}{$\downarrow$}}

\theoremstyle{thmstyleone}

\theoremstyle{thmstyletwo}

\theoremstyle{thmstylethree}

\begin{document}

\title[Adaptive Algorithmic Control]{Beyond Hardware: Adaptive Algorithmic Control by State-Proxy Equalization}

\author{\fnm{Jianlong} \sur{Lu}%
\footnote{Department of Mathematics, National University of Singapore}%
\label{aff:nus}}

\author{\fnm{Hongrui} \sur{Zhang}%
\footnotemark[\value{footnote}]}


\author{\fnm{Thorsten} \sur{Koch}\footnote{Zuse Institute Berlin \& Technische Universität Berlin}}

\author{\fnm{Ying} \sur{Chen}\footnote{Department of Mathematics \& Institute of Operations Research \& Analytics \& Risk Management Institute, National University of Singapore}}

\abstract{Recent advances in quantum computing have been driven primarily by improvements in hardware. Here we show that substantial gains can instead arise from how finite computational resources are allocated throughout a quantum computation. We introduce Adaptive Algorithmic Control (A2C), a software paradigm founded on a State-Proxy Equalization theorem, which proves that the optimal allocation for a state-derived proxy-error functional equalizes cumulative computational hardness rather than physical time. The required computational hardness is inferred directly from the evolving quantum state, avoiding explicit reconstruction of the exponentially large many-body spectrum. Across quantum optimization problems containing up to 156 qubits, combining exact simulations, large-scale supercomputer computations and IBM quantum hardware experiments, A2C improves the low-energy sampling probabilities by 22\% to over 100,000\% under matched circuit depths and measurement budgets.  These results demonstrate that quantum computational performance depends not only on hardware capabilities, but also on how finite computational resources are organized, establishing adaptive algorithmic control as a complementary software pathway for advancing quantum computation.
}


\maketitle

\section{Introduction}
\label{sec:introduction}

The recent progress of quantum computing has been driven primarily by advances in hardware. Larger processors, higher gate fidelities, longer coherence times and the development of quantum error correction continue to expand the scale of quantum computation~\cite{preskill2018quantum}. These advances are indispensable for achieving fault-tolerant quantum computing. At the same time, they have shaped a prevailing view that improved computational performance must primarily come from improved hardware. Yet, as quantum devices become increasingly capable, an equally important question emerges: \emph{how should finite quantum resources be allocated?} This question has long been central to classical computing. The extraordinary performance of modern computers results not only from faster processors, but also from advances in algorithms, compilers and software that determine how computational resources are utilized. Similar opportunities remain largely unexplored in quantum computing.

This distinction matters because quantum computations are dynamically heterogeneous. Along the trajectory of a quantum algorithm, some regions may be traversed with little loss of accuracy, whereas others exhibit rapid state variation, strong many-body correlations or heightened sensitivity to discretization. Allocating the identical computational resources everywhere wastes resources in easy regions while undersampling the regions that determine performance. Consequently, two algorithms executed on the same processor, with the same circuit depth and measurement budget, can produce substantially different outcomes solely because their computational trajectories are organized differently.

Here we introduce \emph{Adaptive Algorithmic Control} (A2C), a software-level framework for allocating computational resources according to the evolving dynamics of a quantum algorithm. Hardware determines which operations are physically available; A2C determines how those operations are distributed along the computation. This is distinct from physical quantum control, which engineers pulses, fields or device-level dynamics. It also differs from schedule-design methods that require spectral gaps or problem-specific analytical information \cite{roland2002quantum,kolodrubetz2017geometry,brady2021optimal,wurtz2022counterdiabaticity}. A2C instead uses state-accessible dynamical information and operates without modifying the underlying hardware. Building on our earlier use of Tx-NQDTs to reconstruct spectral diagnostics and design adaptive quantum-annealing schedules~\cite{Lu2025Transformer},the present framework replaces spectrum-informed scheduling with a state-proxy allocation principle that requires neither explicit spectral reconstruction nor modification of the prescribed operator path.

We establish a general principle, termed \emph{State-Proxy Equalization}, for this allocation problem. For a broad class of discretized quantum evolutions, the optimal trajectory is obtained by equalizing cumulative dynamical difficulty rather than physical time. The relevant difficulty can be estimated from observables of the evolving state, avoiding explicit reconstruction of the many-body spectrum. This result converts adaptive scheduling from an empirical choice into a well-defined optimization principle: computational resolution should be concentrated where the state dynamics indicate that it is most valuable.

We instantiate this principle in quantum optimization. Variational algorithms such as the Quantum Approximate Optimization Algorithm construct solutions through a sequence of parameterized quantum evolutions \cite{farhi2014quantum,zhou2020quantum,cerezo2021variational}. We learn an adaptive control policy from quantum trajectories and use it to redistribute a fixed circuit budget without increasing depth or changing the available gate set. The resulting controller identifies dynamically difficult regions and assigns them greater resolution, while compressing regions in which the state evolves more smoothly. Large-scale control policies are constructed using a learned surrogate of quantum dynamics together with high-performance supercomputing, allowing the approach to be studied beyond the range accessible to exact state-vector propagation.

We evaluate the framework across problem sizes of $N=5$, $20$, $50$, $100$ and $156$ qubits, combining exact calculations, supercomputer-enabled large-scale studies and experiments on IBM quantum processors. Small-system simulations validate the theoretical principle against exact benchmarks, whereas the larger instances assess whether adaptive control remains effective when exact spectral analysis becomes computationally impractical. On quantum hardware, paired comparisons are conducted under identical devices, circuit depths and measurement budgets, isolating the effect of the algorithmic control policy from additional physical resources.

These results show that the performance of a quantum computation is determined not only by the capability of its hardware, but also by how its finite computational budget is allocated. A2C provides a complementary pathway towards improving quantum computation: it does not replace continued advances in quantum hardware, but enables available hardware to be used more effectively. More broadly, our results point towards a hardware--software paradigm for quantum computing, in which advances in quantum processors and adaptive algorithms evolve together on the path towards practical quantum advantage.

\section{Results}

Adaptive Algorithmic Control (A2C) introduces a software-level degree of freedom into quantum computation: the allocation of a fixed computational budget along a quantum trajectory. Rather than increasing circuit depth or hardware resources, A2C redistributes the available computational effort according to the evolving quantum dynamics. We first establish the theoretical principle underlying this adaptive allocation, then show how the required dynamical information can be learned without explicit spectral reconstruction, and finally validate the adaptive schedules across exact simulations, large-scale supercomputer experiments and IBM quantum hardware.

\subsection{State-Proxy Equalization}\label{sec:state_proxy_equalization}

Quantum computations are rarely equally difficult throughout their evolution. We introduce the notion of \emph{computational hardness} $r(s)>0$ to quantify the amount of computational resolution required to accurately evolve the quantum state over a small segment of the quantum trajectory, where $s\in[0,1]$ denotes the normalized trajectory coordinate. A trajectory becomes more difficult whenever the quantum state evolves rapidly, exhibits larger energy fluctuations, or becomes increasingly sensitive to perturbations of the trajectory coordinate. These quantities can all be inferred directly from the evolving quantum state. The explicit construction of $r(s)$ from state-accessible dynamical quantities is presented in the Section Methods.

A larger hardness naturally requires a finer computational resolution. Accordingly, we model the local discretization error by a non-negative \emph{local proxy-error density} $\Phi(s)$, which increases monotonically with the computational hardness. Throughout this work, we choose
$
\Phi(s)=r^2(s),
$
although the theoretical development below applies to any monotone transformation of $r(s)$. Rather than allocating computational resources uniformly along the trajectory, we seek a new computational coordinate $u\in[0,1]$ that redistributes a fixed computational budget according to the \emph{cumulative computational hardness}, namely the accumulated hardness from the beginning of the trajectory to the current point.

\begin{center}
\fbox{
\parbox{0.98\linewidth}{
\paragraph{Theorem 1 (State-Proxy Equalization).}
Let $\Phi(s)\ge0$ denote the local proxy-error density along the quantum trajectory. Among all monotone computational coordinates
$
u:[0,1]\rightarrow[0,1]$ with $
u(0)=0,\;
u(1)=1,
$
the unique coordinate that minimizes the cumulative proxy error is
\begin{equation}
u(s)
=
\frac{\displaystyle\int_0^s\sqrt{\Phi(v)}\,\mathrm{d}v}
{\displaystyle\int_0^1\sqrt{\Phi(v)}\,\mathrm{d}v}.
\label{eq:u}
\end{equation}
}}
\end{center}
The proof follows by minimizing the effective proxy-error functional using the Cauchy–Schwarz inequality and is presented in Section Methods. For the hardness profile introduced above, $\Phi(s)=r^2(s)$, the computational coordinate is simply the normalized cumulative computational hardness, or equivalently,
\begin{equation}
\int_{s_{\ell-1}}^{s_\ell}r(s)\,\mathrm{d}s
=
\frac{1}{p}
\int_0^1r(s)\,\mathrm{d}s,
\qquad
\ell=1,\ldots,p,
\label{eq:equalization}
\end{equation}
where $p$ is the total number of computational intervals. 
Since $r(s)>0$, $u(s)$ is strictly increasing and therefore admits a unique inverse $s=w(u)$, which we refer to as the \emph{adaptive warp}. The adaptive warp maps uniformly spaced computational coordinates back to non-uniform locations along the original quantum trajectory.

\begin{figure*}[t]
\centering
\includegraphics[width=0.86\textwidth]{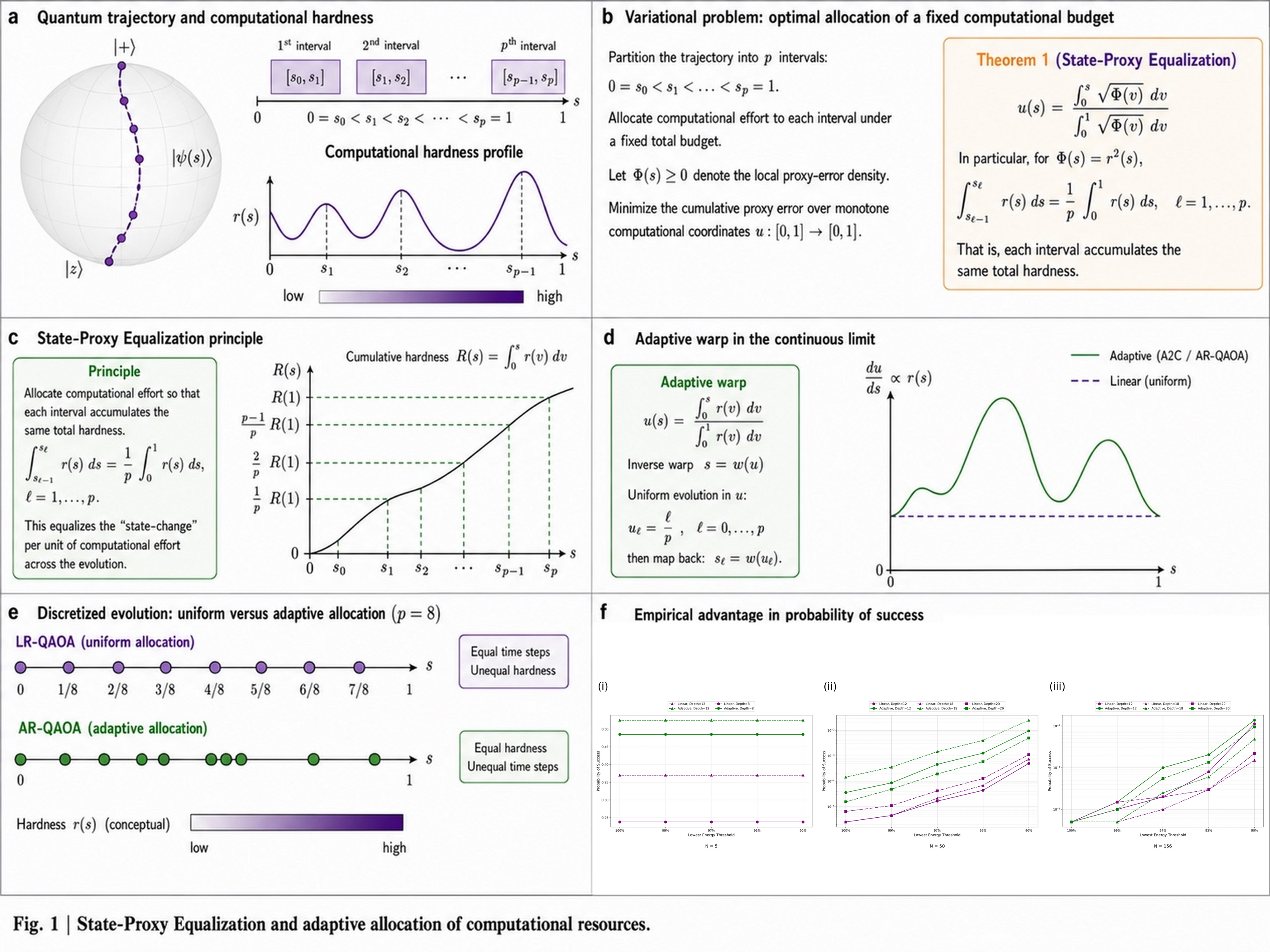}
\caption{
\textbf{State-Proxy Equalization and Adaptive Algorithmic Control (A2C).}
\textbf{a}, A quantum state $|\psi(s)\rangle$ evolves along a normalized trajectory $s\in[0,1]$, where the computational hardness $r(s)$ varies according to the local quantum dynamics. The trajectory is partitioned into $p$ computational intervals $\{[s_{\ell-1},s_\ell]\}_{\ell=1}^{p}$.
\textbf{b}, Variational formulation of the optimal computational allocation. Theorem~1 shows that the optimal computational coordinate is obtained by cumulative state-proxy equalization, leading to equal accumulated hardness in every computational interval.
\textbf{c}, Geometric interpretation of the State-Proxy Equalization principle. Instead of equalizing physical time, the optimal schedule equalizes cumulative computational hardness,
$
\int_{s_{\ell-1}}^{s_\ell} r(s)\,\mathrm{d}s
=
\frac{1}{p}
\int_0^1 r(s)\,\mathrm{d}s,
\qquad
\ell=1,\ldots,p.
$
\textbf{d}, Continuous adaptive warp. The cumulative hardness defines a monotone computational coordinate $u(s)$ and its inverse warp $s=w(u)$, which concentrates computational effort in dynamically difficult regions while preserving the total computational budget.
\textbf{e}, Illustration of discretized evolution. Compared with the uniform linear-ramp schedule (LR-QAOA), A2C reallocates the same number of circuit layers towards regions of higher computational hardness without increasing circuit depth.
\textbf{f}, Empirical improvement in sampling probability. Under identical computational budgets, adaptive scheduling consistently increases the probability of sampling exact and near-optimal solutions for representative quantum optimization problems with $N=5$ (depth $p=8,12$) and $N=50,156$ ($p=12,18,20$) qubits.
}
\label{fig:framework}
\end{figure*}

Theorem~1 establishes a simple but fundamental principle: the optimal allocation of a fixed computational budget does not equalize physical time, but equalizes cumulative computational hardness. The adaptive warp $w$ therefore concentrates computational layers in dynamically difficult regions while preserving the total computational budget. Figure~\ref{fig:framework} summarizes this principle.

\subsection{Learning the adaptive schedule}
\label{sec:learning_schedule}

State-Proxy Equalization determines the optimal resource allocation once the hardness profile $r(s)$ is known. To estimate this profile beyond exact state-vector simulation, we use a Transformer-based Neural Quantum Dynamics Twin (Tx-NQDT) to represent the evolving wavefunction as $\psi_{\phi}(x;s)$, where $x\in\{-1,+1\}^{N}$ is a computational-basis configuration, $s\in[0,1]$ is the normalized trajectory coordinate, and $\phi$ denotes the learned parameters. The model is trained along the reference linear-ramp trajectory and can subsequently be queried at any $s$ without explicitly storing all $2^N$ amplitudes.

For the Hamiltonian $H(s)$ active at coordinate $s$, we define the local energy and trajectory score as
\begin{equation}
E_{\mathrm{loc}}(x;s)
=
\frac{[H(s)\psi_{\phi}](x;s)}{\psi_{\phi}(x;s)},
\qquad
O_s(x;s)=\partial_s\log\psi_{\phi}(x;s).
\label{eq:local_diagnostics}
\end{equation}
Sampling from $\pi_{\phi}(x;s)\propto|\psi_{\phi}(x;s)|^2$, the local energy variance
$\widehat V_H(s)=\operatorname{Var}_{\pi_{\phi}}[E_{\mathrm{loc}}(x;s)]$
measures the energy dispersion of the evolving state, whereas
$\widehat G_{ss}(s)=\operatorname{Var}_{\pi_{\phi}}[O_s(x;s)]$
measures its sensitivity to motion along the trajectory. We combine them into the positive hardness profile
\begin{equation}
r(s)
=
\epsilon
+a\sqrt{\max\{\widehat V_H(s),0\}}
+b\sqrt{\max\{\widehat G_{ss}(s),0\}},
\qquad \epsilon>0,
\label{eq:learned_hardness}
\end{equation}
where $a$ and $b$ control the relative contributions of energy dispersion and trajectory sensitivity, and $\epsilon$ ensures strict positivity. The architecture, training procedure and sampling estimators are described in Section Methods.

Section~\ref{sec:state_proxy_equalization} converts $r(s)$ into the adaptive coordinate $u(s)$ and its inverse warp $s=w(u)$. For a depth-$p$ linear-ramp QAOA schedule,
\begin{equation}
\gamma_{\ell}^{\mathrm{L}}
=
\Delta_{\gamma}\frac{\ell}{p},
\qquad
\beta_{\ell}^{\mathrm{L}}
=
\Delta_{\beta}
\left(1-\frac{\ell-1}{p}\right),
\end{equation}
whereas A2C replaces the uniform coordinate by the learned warp:
\begin{equation}
\gamma_{\ell}^{\mathrm{A}}
=
\Delta_{\gamma}w\!\left(\frac{\ell}{p}\right),
\qquad
\beta_{\ell}^{\mathrm{A}}
=
\Delta_{\beta}
\left[
1-w\!\left(\frac{\ell-1}{p}\right)
\right],
\qquad \ell=1,\ldots,p.
\label{eq:adaptive_ramp}
\end{equation}
Thus, the linear and adaptive schedules retain the matched depth, endpoints and alternating cost--mixer structure. They differ only in the placement of the fixed layer budget: the linear ramp distributes layers uniformly, whereas A2C concentrates them in regions where the learned dynamics indicate greater computational hardness.

\subsection{Adaptive quantum optimization across scales}
\label{sec:adaptive_optimization}

We evaluate whether A2C improves quantum optimization through exact simulations ($N=5,20$), large-scale Tx-NQDT computations on the NVIDIA GH200-based \emph{Lisa} supercomputer at the Zuse Institute Berlin ($N=50,100,156$), and experiments on the IBM Quantum \textit{ibm\_pittsburgh} superconducting quantum processor under a strictly matched
computational budget. For every $(N,p)$ pair, linear-ramp QAOA (LR-QAOA) and
adaptive-ramp QAOA (AR-QAOA) use the matched problem instance, circuit depth,
ramp endpoints, quantum device and measurement budget; only the placement of
the circuit layers along the trajectory is changed. Performance is measured by
the threshold success probability
$
P_{\rm succ}(\rho)
=
\sum_{x:\,C(x)\leq \rho E_{\min}}p(x),$ with $
\rho\in\{1.00,0.99,0.97,0.95,0.90\},$
where $E_{\min}$ is the reference minimum energy. The $100\%$ threshold measures
the probability of sampling a reference optimum, whereas the lower thresholds
measure the probability mass assigned to progressively broader near-optimal
regions. We additionally report the lowest sampled energy, which tests whether
the adaptive schedule discovers a better individual solution even when exact
optimum samples are too rare for reliable probability estimation.

For the small systems, $N=5$ and $N=20$, the optimum is available exactly and
therefore provides a direct benchmark. AR-QAOA substantially increases the
exact-optimum probability at both tested depths. At $N=5$, the gain is
approximately $108$--$109\%$, while at $N=20$ it reaches $319\%$ at $p=8$ and
$673\%$ at $p=12$. Here, gain denotes
the percentage increase over the baseline (thus, a $108\%$ gain corresponds
to $208\%$ of the baseline performance). The improvement persists across the near-optimal thresholds,
showing that A2C redistributes probability mass throughout the low-energy tail
rather than enhancing only a single outcome. The gains become smaller at the
$90\%$ threshold because this region is already broad: for $N=5$, both schedules
place nearly all probability mass within it, leaving little room for further
improvement. The identical lowest energies confirm that, in these exactly
benchmarked cases, the advantage arises from sampling high-quality solutions
more frequently rather than from changing the known optimum.

The larger systems test whether this advantage survives when exact state-vector
propagation and spectral analysis are no longer practical. For $N=50$, AR-QAOA
improves every reported success threshold at every tested depth. The most
pronounced case is $(N,p)=(50,18)$, where the exact and near-optimal success probabilities increase by 231- to 1,084-fold, including an 868-fold increase in reference optimum sampling, and the lowest sampled
energy also improves. At $N=100$ and $N=156$, the strict $100\%$ probability is
largely fixed at the measurement-resolution floor, making near-optimal
probabilities more informative. AR-QAOA improves the $95\%$ and
$90\%$ success probabilities in all six experiments, with gains ranging from
$22\%$ to $350\%$. It also improves the lowest sampled energy for
$(100,12)$, $(100,18)$ and $(156,20)$. These results show that the adaptive
schedule remains effective at scales for which the dynamical hardness cannot be
validated through exact spectral reconstruction.

Increasing the circuit depth does not, by itself, produce monotonic improvement.
The LR baseline can weaken as $p$ increases because deeper circuits accumulate
additional routing and hardware errors, and the strongest relative gain occurs
at $N=50$, $p=18$, where LR-QAOA assigns exceptionally little probability to
the low-energy region. A2C partly compensates for this limitation by placing the
same number of layers at more informative locations, but it does not eliminate
hardware noise: several large-system cases improve the near-optimal probability
mass without improving the single lowest sampled energy. Taken together,
Table~\ref{tab:all_results} shows that A2C most consistently improves the
distribution of sampled solutions, with its principal large-scale advantage
appearing in the near-optimal low-energy tail rather than solely through rare
exact-optimum events.

\begin{table*}[t]
\centering
\caption{
\textbf{Linear-ramp versus adaptive-ramp QAOA across system sizes and circuit depths.}
For each $(N,p)$ case, LR denotes the linear-ramp baseline and AR the adaptive-ramp
result under matched problem instance, circuit depth, ramp endpoints, hardware and
measurement budget. Success probabilities are reported for the reference optimum
($100\%$) and progressively broader near-optimal energy thresholds ($99\%$, $97\%$,
$95\%$ and $90\%$). Gains of AR over LR are shown in parentheses, defined as
$(P_{\mathrm{AR}}/P_{\mathrm{LR}}-1)\times100\%$; thus, a $100\%$ gain corresponds
to twice the LR probability. Green and red indicate positive and negative gains,
respectively. Powers of ten in square brackets apply to the corresponding probability
entries until the next horizontal divider. The final column reports the lowest sampled
energy, where lower values indicate better solutions.
}
\label{tab:all_results}

\begingroup
\setlength{\tabcolsep}{3.2pt}
\renewcommand{\arraystretch}{1.10}
\footnotesize

\begin{tabular}{@{}lccccc c@{}}
\toprule
&
\multicolumn{5}{c}{\textbf{Threshold Success Probability}}
&
\\[-1.5pt]
\cmidrule(lr){2-6}
\textbf{Case}
& \textbf{100\%}
& \textbf{99\%}
& \textbf{97\%}
& \textbf{95\%}
& \textbf{90\%}
& \makecell{\textbf{Lowest}\\\textbf{energy}} \\
\midrule

$(5,8)$-LR
& 0.176
& 0.320
& 0.659
& 0.814
& 1.000
& $-14.654$ \\

$(5,8)$-AR
& 0.367 \gain{108\%}
& 0.505 \gain{58\%}
& 0.772 \gain{17\%}
& 0.916 \gain{13\%}
& 1.000 \loss{0.02\%}
& $-14.654$ $(=)$ \\

$(5,12)$-LR
& 0.408
& 0.569
& 0.825
& 0.942
& 1.000
& $-14.654$ \\

$(5,12)$-AR
& 0.853 \gain{109\%}
& 0.902 \gain{59\%}
& 0.967 \gain{17\%}
& 0.992 \gain{5\%}
& 1.000 \loss{0.01\%}
& $-14.654$ $(=)$ \\

\midrule

$(20,8)$-LR
& $1.88[10^{-4}]$
& $4.94[10^{-4}]$
& $2.82[10^{-3}]$
& $1.02[10^{-2}]$
& $8.05[10^{-2}]$
& $-65.254$ \\

$(20,8)$-AR
& 7.88 \gain{319\%}
& 19.6 \gain{297\%}
& 8.99 \gain{218\%}
& 2.70 \gain{165\%}
& 14.8 \gain{83\%}
& $-65.254$ $(=)$ \\

$(20,12)$-LR
& 3.06
& 7.06
& 3.98
& 1.36
& 9.36
& $-65.254$ \\

$(20,12)$-AR
& 23.6 \gain{673\%}
& 48.1 \gain{582\%}
& 19.9 \gain{402\%}
& 5.26 \gain{287\%}
& 22.0 \gain{135\%}
& $-65.254$ $(=)$ \\

\midrule

$(50,12)$-LR
& $0.500[10^{-6}]$
& $1.17[10^{-6}]$
& $0.517[10^{-5}]$
& $1.50[10^{-5}]$
& $1.57[10^{-4}]$
& $-142.791$ \\

$(50,12)$-AR
& 3.00 \gain{500\%}
& 6.50 \gain{456\%}
& 2.75 \gain{432\%}
& 7.20 \gain{380\%}
& 6.99 \gain{345\%}
& $-142.791$ $(=)$ \\

$(50,18)$-LR
& 0.167
& 0.333
& 0.200
& 0.733
& 1.09
& $-142.748$ \\

$(50,18)$-AR
& 145 \gain{86,726\%}
& 361 \gain{108,308\%}
& 145 \gain{72,400\%}
& 405 \gain{55,152\%}
& 252 \gain{23,019\%}
& $-142.791$ \egain \\

$(50,20)$-LR
& 6.50
& 11.0
& 4.20
& 12.8
& 11.1
& $-142.791$ \\

$(50,20)$-AR
& 15.5 \gain{139\%}
& 48.5 \gain{341\%}
& 19.4 \gain{362\%}
& 58.7 \gain{359\%}
& 50.4 \gain{354\%}
& $-142.791$ $(=)$ \\

\midrule

$(100,12)$-LR
& $0.500[10^{-6}]$
& $1.00[10^{-6}]$
& $0.250[10^{-5}]$
& $0.600[10^{-5}]$
& $0.780[10^{-4}]$
& $-240.023$ \\

$(100,12)$-AR
& 0.500 (\nochange)
& 1.00 (\nochange)
& 0.250 (\nochange)
& 1.25 \gain{108\%}
& 1.41 \gain{81\%}
& $-261.684$ \egain \\

$(100,18)$-LR
& 0.500
& 0.500
& 0.050
& 0.200
& 0.165
& $-230.624$ \\

$(100,18)$-AR
& 0.500 (\nochange)
& 1.00 \gain{100\%}
& 0.150 \gain{200\%}
& 0.300 \gain{50\%}
& 0.390 \gain{136\%}
& $-248.945$ \egain \\

$(100,20)$-LR
& 0.500
& 0.500
& 0.150
& 0.300
& 0.360
& $-239.832$ \\

$(100,20)$-AR
& 0.500 (\nochange)
& 1.00 \gain{100\%}
& 0.300 \gain{100\%}
& 0.750 \gain{150\%}
& 0.650 \gain{81\%}
& $-230.471$ \eloss \\

\midrule

$(156,12)$-LR
& $0.500[10^{-6}]$
& $1.00[10^{-6}]$
& $0.200[10^{-5}]$
& $0.800[10^{-5}]$
& $1.13[10^{-4}]$
& $-323.194$ \\

$(156,12)$-AR
& 0.500 (\nochange)
& 1.50 \gain{50\%}
& 1.00 \gain{400\%}
& 2.05 \gain{156\%}
& 1.38 \gain{22\%}
& $-316.311$ \eloss \\

$(156,18)$-LR
& 0.500
& 0.500
& 0.100
& 0.300
& 0.150
& $-304.358$ \\

$(156,18)$-AR
& 0.500 (\nochange)
& 0.500 (\nochange)
& 0.250 \gain{150\%}
& 0.600 \gain{100\%}
& 0.485 \gain{223\%}
& $-293.684$ \eloss \\

$(156,20)$-LR
& 0.500
& 1.50
& 0.200
& 0.300
& 0.220
& $-301.602$ \\

$(156,20)$-AR
& 0.500 (\nochange)
& 1.00 \loss{33\%}
& 0.550 \gain{175\%}
& 1.35 \gain{350\%}
& 0.955 \gain{334\%}
& $-306.483$ \egain \\

\bottomrule
\end{tabular}

\endgroup
\end{table*}

\subsection{Experimental Insights}
\label{sec:insights}

The experiments consistently support the central premise of A2C. Across the tested system sizes and circuit depths, adaptive allocation consistently improves the principal low-energy probability metrics under matched logical quantum budgets. LR-QAOA and AR-QAOA use the same problem instances, circuit depths, ramp endpoints, hardware platforms and measurement budgets, isolating the organization of the fixed layer budget as the controlled algorithmic difference. 

Two observations are particularly noteworthy. First, the advantage of A2C persists from exactly solvable small systems to large-scale optimization problems beyond the reach of exact state-vector simulation, showing that the approach remains computationally and experimentally viable on the reported instances up to $N=156$, well beyond the range used for exact state-vector validation. Second, the improvement is consistently larger for exact and near-optimal solutions than for broader energy thresholds. As the threshold becomes less restrictive, both algorithms already place substantial probability mass in the corresponding energy region, leaving less room for improvement. A2C therefore primarily shifts probability mass towards the most valuable low-energy solutions rather than uniformly increasing sampling probabilities.

Taken together, these results are consistent with the central prediction of State-Proxy Equalization: redistributing a fixed layer budget according to state-derived dynamical structure can substantially improve low-energy sampling. Across the reported experiments, the organization of a matched logical quantum budget changes solution probabilities by up to orders of magnitude, showing that quantum optimization performance depends not only on available hardware resources but also on how those resources are algorithmically allocated. More broadly, state-informed algorithmic control provides a complementary software pathway for advancing practical quantum computation.

\section*{Acknowledgments}
The authors gratefully acknowledge the computing time made available to them on the high-performance computer ``Lise'' at the NHR center NHR@ZIB. This center is jointly supported by the Federal Ministry of Research, Technology, and Space and the state governments participating in the NHR (www.nhr-verein.de). The authors acknowledge Vishal Sharathchandra Bajpe from IBM’s Quantum Algorithm Engineering team for insightful discussions and assistance with the experimental implementation on IBM Quantum hardware. The authors also thank Dr Stefan Woerner for insightful discussions.

\begingroup
\setlength{\bibsep}{0pt}
\setlength{\parskip}{0pt}
\bibliography{sn-bibliography}
\endgroup

\clearpage

\section{Methods}

Adaptive Algorithmic Control (A2C) comprises four steps: deriving the
fixed-budget equalization rule, estimating a state-dependent dynamical-hardness
profile, constructing a monotone adaptive warp, and comparing linear-ramp and
adaptive-ramp QAOA under matched computational and experimental resources. These sections provide the methodological basis for the theoretical and experimental results in the main text; additional derivations, implementation details, validation analyses and reproducibility information are provided in the Supplementary Information.

\subsection{State-Proxy Equalization}
This section derives the State-Proxy Equalization principle underlying Theorem~1 and Fig.~1; further derivations and theoretical details are provided in Supplementary Information.

Let $s=w(u)$ be an absolutely continuous, strictly increasing map between a
normalized computational coordinate $u\in[0,1]$ and a trajectory coordinate
$s\in[0,1]$, with $w(0)=0$ and $w(1)=1$. For a positive local proxy-error
density $\Phi(s)$, we define
\begin{equation}
\mathcal{E}[w]
=
\int_0^1
\Phi\!\left(w(u)\right)\left[w'(u)\right]^2\,\mathrm{d}u .
\label{eq:proxy_functional}
\end{equation}
This functional quantifies the relative cost of a finite discretization along
the trajectory; it is not assumed to equal QAOA infidelity or hardware success
probability.

Writing $q(s)=du/ds$, with $q(s)>0$ and
$\int_0^1q(s)\,\mathrm{d}s=1$, gives
$\mathcal{E}[q]=\int_0^1\Phi(s)/q(s)\,\mathrm{d}s$. The
Cauchy--Schwarz inequality yields
\begin{equation}
\mathcal{E}[q]
\geq
\left[\int_0^1\sqrt{\Phi(s)}\,\mathrm{d}s\right]^2,
\qquad
q^\star(s)
=
\frac{\sqrt{\Phi(s)}}
{\int_0^1\sqrt{\Phi(v)}\,\mathrm{d}v},
\label{eq:cs_bound}
\end{equation}
where equality uniquely determines the minimizing density up to sets of
measure zero. The corresponding computational coordinate is
\begin{equation}
u^\star(s)
=
\frac{\int_0^s\sqrt{\Phi(v)}\,\mathrm{d}v}
{\int_0^1\sqrt{\Phi(v)}\,\mathrm{d}v}.
\label{eq:equalization_theorem}
\end{equation}

For the implementation used here, we set
$\Phi_{\mathrm{proxy}}(s)=r^2(s)$, where $r(s)>0$ is the
state-derived hardness defined below. Equation~\eqref{eq:equalization_theorem}
then becomes
\begin{equation}
u(s)
=
\frac{\int_0^s r(v)\,\mathrm{d}v}
{\int_0^1 r(v)\,\mathrm{d}v},
\qquad
\int_{s_{\ell-1}}^{s_\ell}r(s)\,\mathrm{d}s
=
\frac{1}{p}\int_0^1r(s)\,\mathrm{d}s .
\label{eq:u_hardness}
\end{equation}
Thus, each of the $p$ computational intervals accumulates the same
hardness. The ideal equalized coordinate satisfies
\begin{equation}
\mathcal{E}_{\mathrm{AR,ideal}}^{\mathrm{proxy}}
=
\left[\int_0^1r(s)\,\mathrm{d}s\right]^2
\leq
\int_0^1r^2(s)\,\mathrm{d}s
=
\mathcal{E}_{\mathrm{LR}}^{\mathrm{proxy}},
\label{eq:E_ar}
\end{equation}
with equality only for an almost-everywhere constant hardness profile.

\subsection{Dynamical-hardness profile}
This section specifies the state-derived hardness profile used to construct the adaptive schedules in the main text; additional estimator and robustness details are provided in Supplementary Information.

We distinguish the normalized accumulated-time coordinate $\tau\in[0,1]$, which parameterizes the piecewise cost–mixer dynamics of the reference LR-QAOA trajectory, from the ramp coordinate $s\in[0,1]$ used to place the QAOA layers. The two coordinates are related by a monotone map $\tau=F(s)$, defined by matching the corresponding cost–mixer segment boundaries. State-derived quantities are first evaluated along $\tau$ and then pulled back through $F$ to obtain the ramp-coordinate hardness profile $r(s)$ used for adaptive schedule construction.

Let $\psi_\phi(x;\tau)$ denote the learned complex amplitude along the reference trajectory, and let $H_k$ be the Hamiltonian active in the segment containing $\tau$. We define the local energy $E_{\mathrm{loc}}(x;\tau)=[H_k\psi_\phi](x;\tau)/\psi_\phi(x;\tau)$, the sampling distribution $\pi_\phi(x;\tau)\propto |\psi_\phi(x;\tau)|^2$, and the trajectory score $O_\tau(x;\tau)=\partial_\tau\log\psi_\phi(x;\tau)$. These quantities define a raw trajectory hardness $r_\tau(\tau)$, which is transferred to the ramp coordinate as $r(s)=F'(s)r_\tau[F(s)]$. The energy and
score variances are
\begin{align}
\widehat V_H(s)
&=
\mathbb E_{\pi_\phi}
\left[
\left|
E_{\mathrm{loc}}
-
\mathbb E_{\pi_\phi}[E_{\mathrm{loc}}]
\right|^2
\right],
\nonumber\\
\widehat G_{ss}(s)
&=
\mathbb E_{\pi_\phi}
\left[
\left|
O_s-\mathbb E_{\pi_\phi}[O_s]
\right|^2
\right],
\nonumber\\
r(s)
&=
\epsilon
+a\sqrt{\max\{\widehat V_H(s),0\}}
+b\sqrt{\max\{\widehat G_{ss}(s),0\}} .
\label{eq:hardness}
\end{align}
For exact evolution within each piecewise-constant segment, the centered trajectory-score variance is proportional to the local-energy variance. We nevertheless retain both estimators in the operational hardness profile because they provide complementary numerical diagnostics of the learned trajectory, rather than representing independent physical hardness components.
Mean subtraction in $\widehat G_{ss}$ removes the contribution of an
$s$-dependent global phase. We used $a=1$, $b=0.2$ and
$\epsilon=10^{-8}$, fixed before the quantum-hardware evaluation. The offset
ensures that the cumulative coordinate is invertible. Only the relative
variation of $r(s)$ affects the warp, because multiplication of the complete
profile by a positive constant leaves Eq.~\eqref{eq:u_hardness} unchanged.
The profile is a dynamical state proxy and is not interpreted as an estimator
of the many-body spectral gap.

\subsection{Neural quantum dynamics twin}
This section describes how Tx-NQDT provides the state information required to estimate the hardness profile beyond exact state-vector simulation; architecture, training and validation details are provided in Supplementary Information.

The Transformer Neural Quantum Dynamics Twin (Tx-NQDT) approximates the
intermediate state along the piecewise cost--mixer trajectory of LR-QAOA
without storing all $2^N$ amplitudes. The cost and mixer Hamiltonians are
$H_C=\sum_i h_iZ_i+\sum_{i<j}J_{ij}Z_iZ_j$ and
$H_M=-\sum_iX_i$, respectively. A depth-$p$ LR-QAOA circuit uses
\begin{equation}
\gamma_\ell^{\mathrm{LR}}
=
\Delta_\gamma\frac{\ell}{p},
\qquad
\beta_\ell^{\mathrm{LR}}
=
\Delta_\beta
\left(1-\frac{\ell-1}{p}\right),
\qquad
\ell=1,\ldots,p .
\label{eq:lr_angles}
\end{equation}
Each layer is decomposed into a cost segment followed by a mixer segment. We
write
$H_k=-(A_k/2)\sum_iX_i+(B_k/2)H_C$, with
$(A_k,B_k)=(0,1)$ for cost segments and $(1,0)$ for mixer segments. Their
durations are $\Delta t_{2\ell-2}=2\gamma_\ell$ and
$\Delta t_{2\ell-1}=2\beta_\ell$, and the normalized segment boundaries are
$s_k=t_k/T$, where $t_k=\sum_{j<k}\Delta t_j$.

For a configuration $x$, the Hamiltonian action is evaluated locally as
\begin{equation}
(H_k\psi_\phi)(x;s)
=
-\frac{A_k}{2}\sum_{i=1}^N\psi_\phi(x^{\oplus i};s)
+
\frac{B_k}{2}E_z(x)\psi_\phi(x;s),
\label{eq:local_h_action}
\end{equation}
where $x^{\oplus i}$ differs from $x$ by one spin flip and
$E_z(x)=\sum_i h_ix_i+\sum_{i<j}J_{ij}x_ix_j$. Evaluation therefore requires
the queried configuration and its $N$ single-spin-flip neighbours rather
than an explicit state vector.

The network represents the rescaled amplitude
$\psi_\phi(x;s)\simeq2^{N/2}\langle x|\psi(s)\rangle
=\exp[\alpha_\phi(x;s)+i\vartheta_\phi(x;s)]$, for which
$\psi_\phi(x;0)=1$. Spin and positional embeddings are processed by a
Transformer with graph-aware attention, while the trajectory coordinate is
introduced through feature-wise linear modulation.

Training minimizes a segment-wise Crank--Nicolson residual. For the
piecewise-constant Hamiltonian $H_k$,
\begin{align}
\mathcal R_k(x)
={}&
\psi_\phi(x;s_{k+1})
+\frac{i\Delta t_k}{2}(H_k\psi_\phi)(x;s_{k+1})
-\psi_\phi(x;s_k)
+\frac{i\Delta t_k}{2}(H_k\psi_\phi)(x;s_k),
\nonumber\\
\mathcal L_k
={}&
\mathbb E_{x\sim q_k}|\mathcal R_k(x)|^2
+\lambda_{\mathrm{norm}}\mathcal L_{\mathrm{norm},k}
+\mathbf 1_{k=0}\lambda_{\mathrm{init}}\mathcal L_{\mathrm{init}},
\label{eq:training_loss}
\end{align}
where
$
\mathcal L_{\mathrm{norm},k}
=
\left(\mathbb E_{q_k}|\psi_\phi(x;s_k)|^2-1\right)^2
+
\left(\mathbb E_{q_k}|\psi_\phi(x;s_{k+1})|^2-1\right)^2
$
and
$\mathcal L_{\mathrm{init}}
=\mathbb E_{q_0}|\psi_\phi(x;0)-1|^2$.
For large systems, $q_k$ was approximated by independent uniform
minibatches, and the conditioned network was optimized sequentially along the
trajectory.

The large-system Tx-NQDT used a six-layer Transformer with graph-aware attention and FiLM trajectory conditioning. Full architecture, training hyperparameters and implementation details are provided in Supplementary Information. After training, configurations were sampled from $\pi_\phi(x;s)$ by
Metropolis--Hastings updates. Local energies and trajectory derivatives were
evaluated on a grid of $s$ values, with automatic differentiation used for
$\partial_s\log\psi_\phi$. These estimates determine
$\widehat V_H(s)$, $\widehat G_{ss}(s)$, and the hardness profile in
Eq.~\eqref{eq:hardness}.

\subsection{Adaptive schedule}
This section converts the learned hardness profile into the finite-depth adaptive schedules evaluated in the main text; additional optimization and sensitivity analyses are provided in Supplementary Information.

Let $R(s)=\int_0^s r(v)\,\mathrm{d}v$. The initial equalized warp is
$w_0(u)=R^{-1}[uR(1)]$, evaluated by trapezoidal quadrature, with initial
nodes $s_j=w_0(j/p)$. The continuum equalization rule therefore provides the initial warp. At finite depth, we refine this warp using a predeclared proxy-only optimization, while keeping the hardness profile fixed and without using quantum-hardware outcomes. For this refinement, the monotone warp is parameterized by
positive increments $\delta_j=s_j-s_{j-1}$, constrained by
$\sum_j\delta_j=1$, using a softmax parameterization with a minimum interval
fraction.

The finite-depth schedule minimizes
\begin{equation}
\mathcal J_{\mathrm{warp}}
=
\sum_{j=0}^{p-1}
\Phi_{\mathrm{proxy}}(s_j^{\mathrm{mid}})
(\Delta s_j)^2
+
\lambda_{\mathrm{sm}}\Omega_{\mathrm{sm}}
+
\lambda_{\mathrm{LR}}\Omega_{\mathrm{LR}},
\label{eq:warp_objective}
\end{equation}
where $s_j^{\mathrm{mid}}$ is the interval midpoint,
$\Omega_{\mathrm{sm}}$ penalizes nonsmooth schedules, and
$\Omega_{\mathrm{LR}}$ penalizes unnecessary displacement from the identity
warp. The cross-entropy optimization used 50 iterations, a population of 96,
12 elites, smoothing parameter 0.7, initial logit standard deviation 0.8,
minimum layer fraction 0.02, $\lambda_{\mathrm{sm}}=0.02$, and
$\lambda_{\mathrm{LR}}=0.005$. The hardness profile remained fixed, and no
quantum-hardware measurements entered the construction.

The final angles are
\begin{equation}
\gamma_\ell^{\mathrm{AR}}
=
\Delta_\gamma
w\!\left(\frac{\ell}{p}\right),
\qquad
\beta_\ell^{\mathrm{AR}}
=
\Delta_\beta
\left[
1-w\!\left(\frac{\ell-1}{p}\right)
\right].
\label{eq:ar_angles}
\end{equation}
The identity warp recovers LR-QAOA. LR and AR therefore have the same qubit
count, logical depth, alternating operator structure and ramp endpoints, and
differ only through a shared monotone deformation of the layer coordinate.

\subsection{Problem instances}
This section defines the benchmark instances underlying the numerical and hardware comparisons in Table~1; complete instance specifications are provided in Supplementary Information.

We consider sparse unconstrained quadratic binary problems
\begin{equation}
C(x)
=
\sum_iQ_{ii}x_i+\sum_{i<j}Q_{ij}x_ix_j,
\qquad
x\in\{0,1\}^N .
\label{eq:qubo}
\end{equation}
Diagonal coefficients were drawn independently from
$\mathrm{Unif}[-11,-2]$. Off-diagonal edges were included independently
with probability $0.18$, with nonzero coefficients drawn from
$\mathrm{Unif}[1,4]$. The complete matrix and pseudorandom seed were retained
for every instance.

Using $x_i=(1+z_i)/2$, the objective becomes
$C(x)=c_0+\sum_i h_iz_i+\sum_{i<j}J_{ij}z_iz_j$, with
$J_{ij}=Q_{ij}/4$ and
$h_i=Q_{ii}/2+\frac14\sum_{j\neq i}Q_{ij}$. Circuits use the corresponding
Ising Hamiltonian, whereas reported objective values are evaluated directly
from Eq.~\eqref{eq:qubo}.

The benchmarks use $N=5,20,50,100$ and $156$, with
$p\in\{8,12\}$ for $N=5,20$ and $p\in\{12,18,20\}$ for the larger
systems. Within every $(N,p)$ comparison, LR and AR use the same QUBO
instance, initial state, circuit depth, ramp scales and evaluation procedure.

\subsection{Computational and hardware execution}
This section details the computational and quantum-hardware protocols used for the matched LR--AR comparisons in the main text; complete execution and compilation settings are provided in Supplementary Information.

Tx-NQDT training and proxy extraction were performed on distributed NVIDIA
A100 GPU resources, with final schedule construction, circuit generation and
postprocessing performed on an NVIDIA GH200 Grace Hopper system. Software,
hardware and run-specific configurations were retained with the computational
artifacts.

Quantum circuits were executed on the IBM Quantum Heron R3
\texttt{ibm\_pittsburgh} superconducting processor (156 qubits) using Qiskit and IBM Quantum Compute SamplerV2. LR and AR members
of a matched pair used the same line-topology routing prescription,
nearest-neighbour cutoff $k=2$, transpiler optimization level 3 and
measurement budget, and were submitted within the same batch whenever
possible. Dynamical decoupling used the XpXm sequence, with Pauli twirling of
gates and measurements. No zero-noise extrapolation or output-level
quasiprobability reconstruction was applied. The adaptive schedule was fixed
before hardware submission.

For archived counts $n_x$, the effective number of shots was determined
directly as $S_{\mathrm{eff}}=\sum_x n_x$, and empirical probabilities were
$\widehat p_{\mathrm{hw}}(x)=n_x/S_{\mathrm{eff}}$. The returned count
denominator, rather than the nominal runtime request, was used for every
circuit.

A re-execution of this benchmark with an upgraded pipeline runs with complete
untruncated cost operators, deterministic SAT-optimized routing with
matched-pair verification, control schedules and error suppression is in
progress and will be reported in an updated version of this manuscript.

\subsection{Performance measures and statistics}
This section defines the success metrics and statistical treatment used for the comparisons in Table~1; additional analyses are provided in Supplementary Information.

Let $C^\star$ denote the independently determined reference minimum. For
$\rho\in(0,1]$, define
\begin{equation}
\mathcal A_\rho
=
\left\{
x:
C(x)-C^\star\leq(1-\rho)|C^\star|
\right\},
\qquad
\widehat P_{\mathrm{succ}}(\rho)
=
\frac{1}{S_{\mathrm{eff}}}
\sum_{x\in\mathcal A_\rho}n_x .
\label{eq:success_prob}
\end{equation}
All reported instances satisfy $C^\star<0$, so that
$\mathcal A_\rho=\{x:C(x)\leq\rho C^\star\}$. We evaluate
$\rho\in\{1.00,0.99,0.97,0.95,0.90\}$; $\rho=1$ selects reference-optimal
solutions, whereas smaller values include progressively broader near-optimal
sets.

The lowest observed objective is
$C_{\mathrm{best}}^{\mathrm{obs}}=\min_{x:n_x>0}C(x)$. When $P_{\mathrm{succ,LR}}(\rho)>0$, the gain of AR over LR is defined as
\[
G(\rho)=
\frac{P_{\mathrm{succ,AR}}(\rho)-P_{\mathrm{succ,LR}}(\rho)}
     {P_{\mathrm{succ,LR}}(\rho)}
\times 100\%.
\]
Thus, a gain of $100\%$ corresponds to an AR success probability twice that of LR.
Because gains can become large near the finite-shot resolution, they are interpreted together with the absolute probabilities.

Threshold counts were treated as finite-shot observations conditional on the
compiled circuit and execution environment. Low-count uncertainty was
represented using count-based binomial intervals rather than symmetric
Gaussian approximations. These intervals describe finite-shot uncertainty
within an execution and do not include variation across independent
calibration windows or compilation seeds.

For every benchmark, the QUBO matrix, random seed, Ising mapping, ramp
parameters, trained-model state, hardness profile, adaptive warp, compiled
circuit metadata and hardware counts were retained. Code, trained models,
benchmark instances and execution metadata will be released with the
published work.

\end{document}


\begin{center}
{\Large\bfseries Supplementary Information for}

\vspace{0.45em}
{\LARGE\bfseries Beyond Hardware: Adaptive Algorithmic Control by State-Proxy Equalization}

\vspace{1.15em}
{\large Jianlong Lu$^{1}$, Hongrui Zhang$^{1}$,
Thorsten Koch$^{2,3}$, and Ying Chen$^{1,4}$}

\vspace{0.9em}
\begin{minipage}{0.94\textwidth}
\small
$^{1}$Department of Mathematics, National University of Singapore, Singapore.

$^{2}$Zuse Institute Berlin, Berlin, Germany.

$^{3}$Technische Universit\"at Berlin, Berlin, Germany.

$^{4}$Institute of Operations Research and Analytics and Risk Management
Institute, National University of Singapore, Singapore.
\end{minipage}

\vspace{0.8em}
\begin{minipage}{0.94\textwidth}
\small
\textit{Correspondence and requests for materials should be addressed to
Ying Chen (matcheny@nus.edu.sg) and Jianlong Lu (jianlong@nus.edu.sg).}
\end{minipage}
\end{center}

\vspace{0.7em}\hrule\vspace{0.6em}

\begingroup
\small
\setlength{\parskip}{0pt}
\tableofcontents
\endgroup

\vspace{0.4em}

\phantomsection

\section*{Guide to the Supplementary Information}
\addcontentsline{toc}{section}{Guide to the Supplementary Information}
\label{supp:guide}

This Supplementary Information develops the theoretical, computational and
experimental foundations of Adaptive Algorithmic Control (A2C). It provides
the derivation of State-Proxy Equalization, the construction and training of
the \TxNQDT{}, the inference of state-derived computational hardness, and the
construction of adaptive finite-depth quantum schedules. It further documents
the numerical validation, large-scale implementation, quantum-hardware
experiments, statistical analysis, robustness studies and reproducibility
procedures supporting the results reported in the main Article.

The Supplementary Notes are organized as follows. Supplementary Note~1
positions A2C relative to prior work in quantum control, QAOA schedule design
and neural quantum-state methods, and delineates the contribution of the
present framework. Supplementary Notes~2--3 introduce the benchmark formulation
and develop the State-Proxy Equalization theory. Supplementary Notes~4--5
describe the \TxNQDT{}, hardness-profile estimation and adaptive schedule
construction. Supplementary Notes~6--10 present numerical validation,
large-scale computational implementation, quantum-hardware evaluation,
statistical analysis, and ablation and robustness studies. Supplementary
Notes~11--12 provide the reproducibility framework and discuss the scope and
generalization of the approach.

\subsection*{Principal notation}

Binary optimization variables are denoted by $b_i\in\{0,1\}$ and spin labels
by $z_i=2b_i-1\in\{-1,+1\}$. The accumulated-time coordinate
$\tau\in[0,1]$ labels the fixed piecewise cost--mixer dynamics, while
$s\in[0,1]$ labels the layer/ramp coordinate and $u\in[0,1]$ is the uniform
allocation coordinate. A declared monotone map $\tau=F(s)$ transfers learned
trajectory quantities to the ramp coordinate; the increasing map $s=w(u)$
then redistributes the $p$ layer coordinates. This distinction separates the learned trajectory coordinate from the layer-allocation coordinate.

\begin{table}[htbp]
\centering
\caption{Principal notation. Only symbols used repeatedly in the
Supplementary Information are listed.}
\label{tab:notation}
\small
\begin{tabularx}{\textwidth}{L{0.19\textwidth}Y}
\toprule
Symbol & Definition \\
\midrule
$N,p$ & Number of binary variables (qubits) and number of QAOA cost--mixer
pairs. \\
$C(b),H_C,H_M$ & QUBO objective, diagonal Ising cost Hamiltonian and
transverse-field mixer. \\
$\gamma_\ell,\beta_\ell$ & Cost and mixer angles in layer $\ell$. \\
$\tau,s,u,F,w$ & Accumulated-time coordinate, ramp coordinate, allocation
coordinate, pullback $\tau=F(s)$ and monotone warp $s=w(u)$. \\
$\psi_\phi(z;\tau),\pi_\phi(z;\tau)$ & Learned complex amplitude and its normalized
Born distribution. \\
$E_{\rm loc},O_\tau$ & Local energy and accumulated-time score
$O_\tau=\partial_\tau\log\psi_\phi$. \\
$\widehat V_H,\widehat G_{\tau\tau}$ & Estimated local-energy variance and
centered trajectory-score variance. \\
$r,\Phi_{\rm proxy}$ & Positive hardness profile and proxy density,
$\Phi_{\rm proxy}=r^2$. \\
$C^\star,A_\rho,P_{\rm succ}$ & Reference minimum, accepted low-energy set and
its sampling probability. \\
$S_{\rm eff}$ & Number of returned shots computed from archived counts. \\
\bottomrule
\end{tabularx}
\end{table}

\section{Relation to prior work}
\label{supp:prior-work}

\subsection{Adaptive Algorithmic Control}

Adaptive Algorithmic Control (\ATwoC{}) is a software paradigm that treats the
organization of finite computational resources as an algorithmic control
variable. Rather than assuming that a prescribed computational budget should
be distributed uniformly according to physical time, circuit layer or another
external coordinate, \ATwoC{} adapts this allocation using information
extracted from the evolving computational state. The underlying computational
path and available resource budget remain prescribed; the control problem is
how that finite budget should be organized along the path.

The central principle developed in this work is State-Proxy Equalization.
Given a positive state-derived measure of computational hardness along a
prescribed path, the associated allocation rule concentrates computational
resolution in more demanding regions and reduces it in less demanding regions
while preserving the total coordinate budget. This separates the amount of
available computational resource from the algorithmic rule used to organize
it, providing a software control layer complementary to improvements in the
underlying hardware. The formal variational statement and its assumptions are
developed in \ref{supp:theory}.

The present work develops and tests this principle through finite-depth quantum
optimization. QAOA provides a concrete realization in which circuit depth
defines a fixed layer budget, the cost--mixer dynamics define a reference
trajectory, and the evolving quantum state supplies information for adaptive
control. The resulting adaptive-ramp QAOA (AR-QAOA) schedule is therefore the
specific realization of \ATwoC{} investigated here, rather than the definition
of \ATwoC{} itself.

\subsection{Control and schedule allocation}

The idea that computational or dynamical resolution should be distributed
nonuniformly has precedents across quantum control and quantum optimization.
Local adiabatic evolution redistributes continuous evolution time using
spectral information, while quantum-geometric and counterdiabatic approaches
use state response or additional controls to suppress nonadiabatic error
\cite{RolandCerf2002,KolodrubetzEtAl2017,SelsPolkovnikov2017,WurtzLove2022}.
Optimal-control analyses connect alternating Hamiltonians with bang--bang and
bang--anneal--bang protocols \cite{YangEtAl2017,BradyEtAl2021}. \ATwoC{}
shares the principle of nonuniform allocation but differs in both the
information used for control and the constrained problem being solved. In the
present realization, the one-dimensional QAOA path, cost and mixer operators,
and number of alternating layers are fixed. The allocation is instead
constructed from a state-accessible hardness profile, without requiring an
instantaneous spectral gap or introducing counterdiabatic operators or
additional layers.

This distinction is particularly clear in comparison with local adiabatic
scheduling. Local adiabatic rules are commonly expressed through an
instantaneous spectral gap and transition matrix element, whereas State-Proxy
Equalization receives a state-derived positive density and determines how
finite computational resolution is distributed along an already prescribed
path. Counterdiabatic and variational gauge-potential methods may enlarge the
available generator set; the present construction does not. Consequently, the
equalization result is an allocation principle for the declared proxy
functional and is not imported as a bound on diabatic excitation.

Within QAOA, structured parameterizations exploit smoothness,
annealing-inspired initializations and parameter transfer
\cite{ZhouEtAl2020,SackSerbyn2021,GaldaEtAl2021,ShaydulinEtAl2023,
SureshbabuEtAl2024,MontanezBarreraMichielsen2025}. The present construction is
likewise low dimensional, but its initialization is generated from an inferred
state trajectory rather than prescribed solely from layer index or transferred
parameters. Fourier, interpolation and transfer strategies could in principle
be combined with this initialization, although such hybrid constructions are
not evaluated here.

Adaptive-QAOA and alternating-operator approaches act on a different control
axis by altering the operator sequence or selecting generators from an
operator pool \cite{ZhuEtAl2022,YanakievEtAl2024,HadfieldEtAl2019}.
The present AR-QAOA realization instead preserves the alternating ansatz and
changes the coordinated cost and mixer angles through a shared monotone warp.
Schedule allocation and operator adaptation are therefore distinct and
potentially complementary algorithmic controls.

\subsection{Neural quantum states as control surrogates}

Neural quantum states provide compact variational representations of some
many-body wavefunctions and support sampled evaluation of amplitudes and local
observables \cite{CarleoTroyer2017,GaoDuan2017,SharirEtAl2020}.
Time-dependent variants approximate real-time evolution
\cite{SchmittHeyl2020,GutierrezMendl2022}, and neural states have been used to
simulate QAOA \cite{MedvidovicCarleo2021}. Transformer wavefunctions provide a
natural mechanism for representing long-range correlations
\cite{ViterittiEtAl2023,ZhangDiVentra2023}.

The \TxNQDT{} uses these developments in a different role: as an offline
control surrogate for a prescribed reference trajectory. The trained model is
not the final sampler in the quantum-hardware experiments. Instead, its learned
amplitudes and trajectory derivatives are converted into state-derived local
statistics, a positive hardness profile, a cumulative coordinate and,
ultimately, an executable quantum schedule. Once this schedule is frozen,
ordinary QAOA circuits are constructed and executed independently of the
surrogate.

This separation between trajectory reconstruction and quantum execution is
central to the present architecture. It allows state information that would be
difficult to obtain directly from repeated hardware measurements to inform the
allocation offline, while leaving the final quantum circuit as a conventional
finite-depth circuit. At the same time, surrogate approximation and sampling
error can propagate into the inferred profile and schedule. Their numerical
validation is therefore treated separately from the subsequent QAOA outcome
evaluation in \ref{supp:validation} and
\ref{supp:statistics}.

\subsection{Positioning of the present framework}

The present framework brings together four elements: (i) an exact allocation
rule for a declared positive proxy functional; (ii) state-derived estimation
of the corresponding hardness profile without explicit full state-vector
storage at large $N$; (iii) conversion of that profile into a shared monotone
warp of a fixed finite-depth quantum path; and (iv) matched evaluation of the
resulting adaptive schedules against the corresponding linear-ramp baseline.
Together, these elements implement a state-informed software control layer
between a prescribed computational path and its finite-resource execution.

In this sense, \ATwoC{} differs from methods whose primary objective is to add
quantum resources, enlarge the operator ansatz, or directly optimize an
unconstrained set of circuit parameters. In the realization studied here, the
logical depth, operator family and ramp endpoints are held fixed, while the
allocation of the existing layer budget is changed. The phrase
``spectrum-free'' therefore means that construction of the adaptive schedule
does not require eigendecomposition or explicit spectral-gap reconstruction.
Likewise, ``fixed depth'' denotes a matched number of logical cost--mixer pairs;
it does not imply equal physical pulse duration or reduced wall-clock runtime.

The broader distinction is between \emph{what computational pathway is
available} and \emph{how finite computational resolution is allocated along
that pathway}. Local adiabatic and counterdiabatic approaches modify the
continuous-time control strategy using spectral, geometric or auxiliary-control
information; structured QAOA methods constrain or transfer circuit parameters;
adaptive-operator methods modify the available generator sequence; and neural
quantum states are commonly used as variational representations or simulators.
\ATwoC{} instead introduces state-informed resource allocation as a distinct
algorithmic control layer. In the realization studied here, the \TxNQDT{}
provides the state-derived control signal, State-Proxy Equalization determines
the allocation principle, and AR-QAOA converts that allocation into an
executable finite-depth quantum schedule.

These control axes are not mutually exclusive: State-Proxy Equalization could
in principle be combined with parameter transfer, adaptive operator selection
or other control families, although such hybrid constructions are beyond the
experiments reported here.

\section{Conventions and benchmark definitions}
\label{supp:conventions}

\subsection{QUBO and Ising conventions}

For $b\in\{0,1\}^N$, the objective and spin transformation are
\begin{align}
C(b)&=\sum_i Q_{ii}b_i+\sum_{i<j}Q_{ij}b_i b_j, & z_i&=2b_i-1.
\label{eq:qubo}
\end{align}
Substitution of $b_i=(1+z_i)/2$ gives
\begin{align}
C(b)&=c_0+\sum_i h_i z_i+\sum_{i<j}J_{ij}z_i z_j,\nonumber\\
J_{ij}&=\frac{Q_{ij}}4,\qquad
h_i=\frac{Q_{ii}}2+\frac14\!\left(\sum_{j<i}Q_{ji}+\sum_{j>i}Q_{ij}\right),\qquad
c_0=\frac12\sum_iQ_{ii}+\frac14\sum_{i<j}Q_{ij}.
\label{eq:ising-map}
\end{align}
The circuit Hamiltonians are $H_C=\sum_i h_iZ_i+\sum_{i<j}J_{ij}Z_iZ_j$ and
$H_M=-\sum_iX_i$.  The constant $c_0$ is omitted from $H_C$ because it produces
only a global phase, but all reported objective values are evaluated from
Eq.~\eqref{eq:qubo}.  Diagonal coefficients are sampled from
$\mathrm{Unif}[-11,-2]$; edges are included independently with probability
$0.18$, and included off-diagonal coefficients are sampled from
$\mathrm{Unif}[1,4]$. The source data contain each reported matrix and its
generation seed.

The parameters were chosen jointly to define a moderately sparse, frustrated
benchmark regime rather than to represent a specific application. The
heterogeneous negative diagonal terms favor variable selection, while positive
pairwise couplings introduce competing penalties at a smaller scale; an edge
probability of $0.18$ preserves sparse interactions while increasing the number
of coupled constraints with system size. The same parameter distributions are
used at every $N$, and each matched LR--AR comparison uses the identical
realized QUBO instance.

\subsection{QAOA order and schedules}

We use the standard alternating cost--mixer construction of QAOA
\cite{FarhiGoldstoneGutmann2014}.

With $U_C(\gamma)=e^{-\ii\gamma H_C}$ and
$U_M(\beta)=e^{-\ii\beta H_M}$, the depth-$p$ state is
\begin{equation}
\ket{\psi_p(\bm\gamma,\bm\beta)}=
U_M(\beta_p)U_C(\gamma_p)\cdots U_M(\beta_1)U_C(\gamma_1)
\ket{+}^{\otimes N}.
\label{eq:qaoa-state}
\end{equation}
The cost unitary acts before the mixer in each layer.  The linear reference and
adaptive schedules are
\begin{align}
\gamma_\ell^{\rm LR}&=\Delta_\gamma\frac{\ell}{p}, &
\beta_\ell^{\rm LR}&=\Delta_\beta\!\left(1-\frac{\ell-1}{p}\right),\nonumber\\
\gamma_\ell^{\rm AR}&=\Delta_\gamma w\!\left(\frac{\ell}{p}\right), &
\beta_\ell^{\rm AR}&=\Delta_\beta\!\left[1-w\!\left(\frac{\ell-1}{p}\right)\right].
\label{eq:lr-ar-schedules}
\end{align}
Here $\Delta_\gamma>0$, $\Delta_\beta>0$, $w(0)=0$, $w(1)=1$ and $w$ is
strictly increasing. Therefore both
schedules have the same depth, operator order and endpoint amplitudes.  No
post-warp rescaling of the angle sums is applied: such a rescaling would in
general destroy the endpoint equality asserted in the main Article.  The fixed
budget in the theorem and experiments is the layer-coordinate/depth budget,
not the sum of rotation angles.

\subsection{Reference trajectory and coordinate pullback}

The \LRQAOA{} circuit is decomposed into $2p$ piecewise-constant segments.  In
the indexing $k=0,\ldots,2p-1$, cost and mixer durations are
$\Delta t_{2\ell-2}=2|\gamma_\ell^{\rm LR}|$ and
$\Delta t_{2\ell-1}=2|\beta_\ell^{\rm LR}|$.  With
$t_k=\sum_{j<k}\Delta t_j$, $T=\sum_j\Delta t_j$ and $\tau_k=t_k/T$, the active
Hamiltonian is
\begin{equation}
H_k=-\frac{A_k}{2}\sum_iX_i+\frac{B_k}{2}H_C,\qquad
(A_k,B_k)=\begin{cases}(0,1),&\text{cost segment},\\(1,0),&\text{mixer segment}.
\end{cases}
\label{eq:piecewise-hamiltonian}
\end{equation}
The coordinate $\tau\in[0,1]$ labels accumulated normalized segment time; it is
not physical annealing time and is not a continuous interpolation with
$A(\tau)=1-\tau$ and $B(\tau)=\tau$. At internal boundaries, quantities
involving $H_k$ or a derivative are evaluated one-sidedly.

The schedule warp acts on a distinct ramp coordinate $s$. To connect the
learned trajectory to that coordinate, introduce uniform logical half-layer
breaks $x_m=m/(2p)$ and the unique continuous piecewise-linear increasing map
$F$ satisfying $F(x_m)=\tau_m$. The even breaks obey $x_{2\ell}=\ell/p$, so the
identity warp places every layer at its LR ramp coordinate. A raw positive
trajectory density $r_\tau(\tau)$ is transferred by
\begin{equation}
r(s)=F'(s)r_\tau[F(s)],\qquad
\int_0^s r(v)\,dv=\int_0^{F(s)}r_\tau(\tau)\,d\tau.
\label{eq:coordinate-pullback}
\end{equation}
Thus equalization is invariant under the declared relabelling and
$w(u)=u$ exactly recovers Eq.~\eqref{eq:lr-ar-schedules}. The half-layer choice
$x_m=m/(2p)$ is part of the algorithm and must be used for both LR and AR; a
different interpolation would define a different schedule constructor. At a
half-layer knot, $F'$ and the pulled density use the following-segment
(right-continuous) value, with the final endpoint assigned to the last segment.
The assumptions $\Delta_\gamma,\Delta_\beta>0$ make every LR segment duration
positive and hence make $F$ strictly increasing.

\subsection{Matched resources and outcome definitions}

\begin{table}[htbp]
\centering
\caption{Matched quantities in each reported \LRQAOA{}--\ARQAOA{} pair.  The
controlled difference is the monotone layer-coordinate warp.}
\label{tab:matched-resources}
\small
\begin{tabularx}{\textwidth}{L{0.33\textwidth}YY}
\toprule
Quantity & \LRQAOA{} & \ARQAOA{} \\
\midrule
QUBO, Ising map and initial state & Identical & Identical \\
Cost--mixer pairs and operator order & $p$; cost then mixer & Same \\
Ramp endpoints & $\Delta_\gamma,\Delta_\beta$ & Same \\
Layer coordinate & Identity & Frozen learned warp \\
Device and compilation prescription & Matched & Matched \\
Requested measurement budget & Matched & Matched \\
Decoding, objective and thresholds & Identical & Identical \\
\bottomrule
\end{tabularx}
\end{table}

The suite uses $N\in\{5,20,50,100,156\}$, with $p\in\{8,12\}$ for the small
systems and selected $p\in\{12,18,20\}$ for the larger systems.  For a
reference value $C^\star<0$, define
\begin{align}
A_\rho&=\{b:C(b)-C^\star\le(1-\rho)|C^\star|\}
=\{b:C(b)\le\rho C^\star\},\nonumber\\
P_{\rm succ}(\rho)&=\sum_{b\in A_\rho}p(b),\qquad
\rho\in\{1.00,0.99,0.97,0.95,0.90\}.
\label{eq:success-def}
\end{align}
At $\rho=1$, all degenerate strings attaining $C^\star$ are included.  The
term ``exact optimum'' is used only when exhaustive enumeration or a solver
certificate establishes global optimality; otherwise ``reference optimum'' is
used.  Because the percentage thresholds depend on the additive origin of the
objective, they are benchmark-specific low-energy sets rather than
additive-shift-invariant approximation ratios.

\subsection{Reference values and benchmark provenance}

Each benchmark record contains the full upper-triangular QUBO matrix,
generation seed, $N$, edge density, coefficient convention and a checksum.
For small instances, exhaustive enumeration supplies both the minimum and its
degeneracy. For larger instances, a value may be labelled globally optimal
only when the solver emits a verifiable certificate with a zero or explicitly
reported optimality gap. Otherwise the lowest independently obtained value
is denoted $C_{\rm ref}$ and all threshold sets are defined relative to that
reference. Replacing $C^\star$ by $C_{\rm ref}$ changes the interpretation:
the resulting statistic is still reproducible, but it is not the probability
of being within a stated percentage of the unknown global optimum.

Three objective summaries are retained because they answer different
questions. The mean $\E[C]$ describes the full returned distribution; the
best observed value is an order statistic whose expectation depends strongly
on the shot count; and $P_{\rm succ}(\rho)$ measures mass in a predeclared
low-energy set. A fair comparison therefore uses matched requested-shot
budgets, reports the returned total for each circuit and treats common-shot
subsampling as a sensitivity analysis. It reports all three summaries without
selecting the most favourable one after inspection. Since $C(b)$ can be shifted by a constant without changing the
unitary up to global phase, the raw QUBO convention and $c_0$ are archived even
though $c_0$ is omitted from circuit generation.

The benchmark suite contains one reported random matrix at each listed size,
not an ensemble large enough to estimate an instance-distribution average.
Size trends are consequently descriptive. Claims of typical-case scaling or
generalization to a random-QUBO population require independently generated
test instances and are outside the scope of the reported benchmark suite.

\section{State-Proxy Equalization}
\label{supp:theory}

\subsection{Variational problem and theorem}

Let $s=w(u)$ be absolutely continuous and nondecreasing, with
$w(0)=0$ and $w(1)=1$. For a measurable density
$\Phi:(0,1)\to[0,\infty)$ with $0<\int_0^1\sqrt{\Phi}<\infty$, define
\begin{equation}
\mathcal E[w]=\int_0^1\Phi(w(u))[w'(u)]^2\,du.
\label{eq:proxy-functional}
\end{equation}
In allocation-density form, $q(s)=du/ds\ge0$ satisfies
$\int_0^1q(s)ds=1$ and
$\mathcal E[q]=\int_0^1\Phi(s)/q(s)\,ds$, with $q>0$ wherever
$\Phi>0$ and the convention $0/0=0$.

\begin{theorem}[State-Proxy Equalization]
The relaxed allocation problem has the unique density up to null sets and
infimum
\begin{equation}
q^\star(s)=\frac{\sqrt{\Phi(s)}}{\int_0^1\sqrt{\Phi(v)}\,dv},
\qquad
\mathcal E_{\min}=\left[\int_0^1\sqrt{\Phi(s)}\,ds\right]^2.
\label{eq:equalization-solution}
\end{equation}
Consequently, $u^\star(s)=\int_0^s q^\star(v)dv$. If $\Phi>0$ almost
everywhere, $w^\star=(u^\star)^{-1}$ is an ordinary inverse and attains the
minimum. Otherwise the formula defines the relaxed optimum and a generalized
monotone inverse; strictly increasing ordinary allocations approach the same
infimum.
\end{theorem}

\begin{proof}
Cauchy--Schwarz and $\int q=1$ give
\begin{equation}
\left(\int_0^1\sqrt{\Phi(s)}\,ds\right)^2
=\left(\int_0^1\sqrt{\frac{\Phi(s)}{q(s)}}\sqrt{q(s)}\,ds\right)^2
\le \int_0^1\frac{\Phi(s)}{q(s)}\,ds.
\end{equation}
Equality holds precisely when $\sqrt{\Phi(s)/q(s)}$ is proportional to
$\sqrt{q(s)}$, which yields Eq.~\eqref{eq:equalization-solution} after
normalization. Strict positivity almost everywhere makes $u^\star$ strictly
increasing. If $\Phi$ vanishes on an interval, $u^\star$ is constant there;
assigning an arbitrarily small positive allocation to the zero set gives a
sequence of ordinary coordinates converging to the displayed infimum.
\end{proof}

The operational profile is strictly positive because of the offset in $r(s)$,
so the implemented cumulative coordinate has an ordinary inverse. This
distinguishes the general nonnegative theorem from the stricter condition used
by the schedule-construction algorithm.

\subsection{Hardness form and equal intervals}

The implementation uses $\Phi_{\rm proxy}(s)=r(s)^2$ with $r(s)>0$.  Hence
\begin{equation}
u^\star(s)=\frac{\int_0^s r(v)\,dv}{\int_0^1r(v)\,dv},
\qquad
\int_{s_{\ell-1}}^{s_\ell}r(s)\,ds
=\frac1p\int_0^1r(s)\,ds,
\label{eq:equal-hardness}
\end{equation}
where $s_\ell=w^\star(\ell/p)$.  The identity warp has
$\mathcal E_{\rm LR}=\int_0^1r(s)^2ds$, while the ideal equalized value is
$\mathcal E_{\rm eq}=(\int_0^1r(s)ds)^2$.  Therefore
\begin{equation}
\mathcal E_{\rm LR}-\mathcal E_{\rm eq}
=\int_0^1[r(s)-\bar r]^2ds,\qquad
1-\frac{\mathcal E_{\rm eq}}{\mathcal E_{\rm LR}}
=\frac{\Var_{s\sim U[0,1]}[r(s)]}{\E_{s\sim U[0,1]}[r(s)^2]}.
\label{eq:proxy-opportunity}
\end{equation}
This identity quantifies the opportunity within the declared proxy: a constant
profile already makes the identity coordinate optimal.  It does not imply a
corresponding improvement in success probability.

\subsection{State quantities and their exact relation}

For a normalized state $\psi(z;\tau)=\braket{z|\psi(\tau)}$, define the Born
distribution $\pi(z;\tau)=|\psi(z;\tau)|^2$, local energy
$E_{\rm loc}=[H(\tau)\psi]/\psi$ and score
$O_\tau=\partial_\tau\log\psi$. With the
complex-variance convention $\Var(Y)=\E|Y-\E Y|^2$,
\begin{align}
V_H(\tau)&=\Var_\pi(E_{\rm loc})
=\braket{H^2}-\braket{H}^2,\nonumber\\
G_{\tau\tau}(\tau)&=\Var_\pi(O_\tau)
=\braket{\partial_\tau\psi|\partial_\tau\psi}
-|\braket{\psi|\partial_\tau\psi}|^2.
\label{eq:state-identities}
\end{align}
The second expression is the Fubini--Study metric along the realized path.
Inside a piecewise-constant segment,
$\partial_\tau\ket{\psi}=-\ii T H_k\ket{\psi}$, and therefore
\begin{equation}
G_{\tau\tau}(\tau)=T^2V_H(\tau).
\label{eq:metric-variance-redundancy}
\end{equation}
Thus the energetic and score terms are not independent physical directions on
an exact segment; they are two estimators of the same projective speed, up to
the fixed coordinate factor $T$. Their joint use in the raw trajectory profile
\begin{equation}
r_\tau(\tau)=\epsilon+a\sqrt{\max(\widehat V_H(\tau),0)}
+b\sqrt{\max(\widehat G_{\tau\tau}(\tau),0)},
\quad (a,b,\epsilon)=(1,0.2,10^{-8}),
\label{eq:hardness-profile}
\end{equation}
is an operational robustness construction, not a combination of two
independent exact observables. The ramp-coordinate density used by the theorem
and schedule is the pullback $r(s)=F'(s)r_\tau[F(s)]$ in
Eq.~\eqref{eq:coordinate-pullback}. Equation
\eqref{eq:metric-variance-redundancy} is also a stringent internal consistency
check on the learned derivative.

\subsection{Invariances and robustness}

If $\widetilde\psi(z;\tau)=c(\tau)e^{\ii\chi(\tau)}\psi(z;\tau)$, then the local-energy
ratio is unchanged and the score shifts by a configuration-independent term;
centering therefore leaves $G_{\tau\tau}$ invariant. Multiplying the complete
hardness profile by a constant also leaves the cumulative coordinate unchanged.

Suppose a positive estimate $\widehat r$ obeys
$(1-\delta)r\le\widehat r\le(1+\delta)r$ with $0\le\delta<1$.  If
$\widehat q=\widehat r/\int\widehat r$, then evaluation with the true proxy
gives
\begin{equation}
\mathcal E_{\min}\le \mathcal E_r[\widehat q]
\le\frac{1+\delta}{1-\delta}\mathcal E_{\min}.
\label{eq:robustness-bound}
\end{equation}
The bound concerns loss of proxy optimality; it is not a bound on state error
or optimization performance.

\subsection{Coordinate covariance and boundary conventions}

The functional is tied to the declared reference coordinate. If a smooth
monotone relabelling $\tau=f(x)$ is introduced, the projective line element
obeys $G_{xx}=G_{\tau\tau}[f'(x)]^2$. The energy variance alone has no
derivative factor. Consequently, profile weights calibrated in one coordinate
cannot be copied to another coordinate and interpreted identically. The raw
profile uses normalized accumulated segment time and is transferred to the
ramp coordinate only by Eq.~\eqref{eq:coordinate-pullback}.

At a switching point the Hamiltonian jumps and the path is continuous but not
differentiable. The left and right limits of $V_H$ and $O_\tau$ can differ. The
implementation stores both one-sided values, performs smoothing only within
open segments, and assigns the exact boundary coordinate to the following
segment when constructing the cumulative integral. Alternative assignments
change a set of measure zero in the continuum theorem, but can affect a coarse
discrete grid; the convention is therefore part of the reproducibility record.

The offset $\epsilon$ has two roles: it prevents flat intervals from making
the cumulative coordinate non-invertible and limits the largest discrete step
when the estimated profile is nearly zero. It is not a physical energy scale.
Sensitivity to $\epsilon$ must be checked before interpreting narrow profile
features. A monotone interpolant is used for both the cumulative function and
its inverse so that numerical overshoot cannot generate a negative layer
spacing.

\subsection{Estimator identities used for validation}

For any normalized exact state and Hermitian $H$, the local-energy identities
$\E_\pi[E_{\rm loc}]=\langle H\rangle$ and
$\E_\pi[|E_{\rm loc}|^2]=\langle H^2\rangle$ follow by inserting the
computational basis. Likewise
$\E_\pi[O_\tau]=\langle\psi|\partial_\tau\psi\rangle$
and centering removes the parallel, gauge-dependent component of the
derivative. These equalities justify sampled estimators but do not make them
unbiased when samples are correlated or generated from an approximate model.

In the exact segment dynamics, define
$D(\tau)=\widehat G_{\tau\tau}(\tau)-T^2\widehat V_H(\tau)$. The mean and scale-normalized
maximum of $D$ are recorded as diagnostic quantities. A discrepancy can arise
from amplitude error, derivative error, Monte Carlo error or using the wrong
one-sided Hamiltonian. It should not be removed by tuning $a/b$, because that
would use an identity violation to calibrate the schedule. Instead, the state
and derivative validation tests in \ref{supp:validation}
must identify the source.

Finally, the Cauchy--Schwarz proof is a continuum statement. At finite grid
resolution the implemented warp minimizes neither an arbitrary discretization
of Eq.~\eqref{eq:proxy-functional} nor the hardware outcome exactly. The grid,
quadrature and interpolation tests quantify convergence to the continuum rule;
the optional finite-dimensional proxy refinement described below is kept
separate from the theorem.

\section{Transformer Neural Quantum Dynamics Twin}
\label{supp:txnqdt}

\subsection{Representation and conditioning}

The \TxNQDT{} represents the rescaled amplitude along the fixed \LRQAOA{}
trajectory,
\begin{equation}
\psi_\phi(z;\tau)\simeq 2^{N/2}\braket{z|\psi_{\rm LR}(\tau)}
=\exp[\alpha_\phi(z;\tau)+\ii\vartheta_\phi(z;\tau)].
\label{eq:txnqdt-amplitude}
\end{equation}
The rescaling makes $\psi_\phi(z;0)=1$ for the initial uniform
superposition.  It cancels from normalized probabilities, local-energy ratios
and centered score variances.  The production parameterization clamps the log
amplitude to $[-12,12]$ and maps the raw phase through
$\vartheta_\phi=\pi\tanh\xi_{\rm raw}$. This numerical choice does not impose
a real-wavefunction gauge.

Each spin is embedded together with a learned site embedding.  A classification
token provides a global channel, and direct spin-to-spin attention is masked to
one-hop neighbours in the QUBO graph. The scalar $\tau$ is introduced by
feature-wise linear modulation: if $e_i$ is a site token, then
$\widetilde e_i(\tau)=g_\phi(\tau)\odot e_i+c_\phi(\tau)$. Six norm-first transformer
blocks with width $128$, four heads and feed-forward width $512$ are used in
the large-system configuration.  The final readout returns a complex
log-amplitude.  This architecture is instance-specific and its parameter count
depends on $N$ through the site embeddings and readout; the present work does
not claim cross-size transfer \cite{VaswaniEtAl2017,PerezEtAl2018}.

\subsection{Attention, output and parameter audit}

For head $h$ in block $m$, masked attention has the standard form
\begin{equation}
\operatorname{Attn}_h(Q,K,V)=
\operatorname{softmax}\!\left(\frac{Q_hK_h^\mathsf{T}}{\sqrt{d_h}}+M\right)V_h,
\qquad
M_{ij}=\begin{cases}0,&i=j\ \text{or}\ (i,j)\in E,\\-\infty,&\text{otherwise},\end{cases}
\label{eq:masked-attention}
\end{equation}
with the classification token connected to every site. Residual connections,
layer normalization and the feed-forward sublayer are applied in norm-first
order. The global token and pooled site features enter separate linear heads
for $\alpha_\phi$ and $\xi_{\rm raw}$. The released configuration records the
number of trainable parameters for every instance; this is required because
site-dependent embeddings make the count vary with $N$.

The one-hop mask is an architectural inductive bias, not a truncation of the
QUBO Hamiltonian. Every nonzero $J_{ij}$ in the archived QUBO remains in the diagonal local
energy. The graph-attention mask changes the neural architecture but does not
sparsify the Hamiltonian. Information can propagate beyond one
hop through stacked attention blocks and the global token. Confusing the mask
with Hamiltonian sparsification would define a different problem, so the edge
list used by the model and the coefficient list used by
Eq.~\eqref{eq:local-h-action} are hashed and checked independently.

\subsection{Local Hamiltonian action}

Let $z^{\oplus i}$ denote $z$ with spin $i$ flipped and
$E_z(z)=\sum_i h_i z_i+\sum_{i<j}J_{ij}z_i z_j$.  In segment $k$,
\begin{equation}
[H_k\psi_\phi](z;\tau)=-\frac{A_k}{2}\sum_{i=1}^{N}
\psi_\phi(z^{\oplus i};\tau)+\frac{B_k}{2}E_z(z)\psi_\phi(z;\tau).
\label{eq:local-h-action}
\end{equation}
Thus a queried configuration requires its $N$ single-spin-flip neighbours,
not all $2^N$ amplitudes.  Neighbour evaluations are processed in chunks of 16
during training and four during hardness extraction.  Chunking changes memory
use and floating-point accumulation order, but not the estimator.

\subsection{Physics-informed training objective}

On segment $k$, the Crank--Nicolson relation \cite{CrankNicolson1947} is
\begin{equation}
(I+\ii\Delta t_kH_k/2)\ket{\psi(\tau_{k+1})}
=(I-\ii\Delta t_kH_k/2)\ket{\psi(\tau_k)}.
\label{eq:crank-nicolson}
\end{equation}
The sampled residual is therefore
\begin{align}
R_k(z)&=\psi_\phi(z;\tau_{k+1})+\frac{\ii\Delta t_k}{2}
[H_k\psi_\phi](z;\tau_{k+1})\nonumber\\
&\quad-\psi_\phi(z;\tau_k)+\frac{\ii\Delta t_k}{2}
[H_k\psi_\phi](z;\tau_k),\nonumber\\
\mathcal L_k&=\E_{z\sim q_k}|R_k(z)|^2
+\lambda_{\rm norm}\mathcal L_{{\rm norm},k}
+\delta_{k0}\lambda_{\rm init}\mathcal L_{\rm init},
\label{eq:training-loss}
\end{align}
where
$\mathcal L_{{\rm norm},k}=(\E_{q_k}|\psi_\phi(\tau_k)|^2-1)^2
+(\E_{q_k}|\psi_\phi(\tau_{k+1})|^2-1)^2$ and
$\mathcal L_{\rm init}=\E_{q_0}|\psi_\phi(z;0)-1|^2$.
For $2^N\le512$, expectations are enumerated; otherwise, independent uniform
minibatches are used.  The rescaled-state identity
$2^{-N}\sum_z|\psi_\phi(z;\tau)|^2=1$ makes the normalization term estimable from
uniform samples.

One coordinate-conditioned model is trained sequentially over the $2p$
segments, carrying model and optimizer state forward. The configured setting
is 80 epochs per segment, nominal global batch size 4096 and Adam learning rate
$5\times10^{-4}$ \cite{KingmaBa2015}. The run record defines the number of minibatches in an
epoch and reports the resulting optimizer-step count; ``epoch'' is not silently
equated with one update. The remaining settings are
$(\lambda_{\rm norm},\lambda_{\rm init})=(10^{-2},10^{-1})$, and gradient-norm
clipping at one.  Distributed ranks use seeds $88+1000r$.  Automatic mixed
precision is disabled; gradient checkpointing is enabled.  Because a shared
model can forget earlier coordinates during sequential training, final
validation queries the complete trajectory rather than only the last
segment.

\begin{table}[htbp]
\centering
\caption{Production \TxNQDT{} configuration.}
\label{tab:txnqdt-config}
\small
\begin{tabularx}{\textwidth}{L{0.43\textwidth}Y}
\toprule
Item & Value \\
\midrule
State representation & Complex log amplitude and phase \\
Transformer & Width 128; 6 blocks; 4 heads; feed-forward width 512 \\
Conditioning and graph structure & FiLM in $\tau$; one-hop graph mask; global
classification token \\
Output safeguards & Log amplitude $[-12,12]$; phase $\pi\tanh\xi$ \\
Training & Adam; $5\times10^{-4}$; 80 epochs per segment; optimizer steps
recorded separately; batch 4096 \\
Loss weights & $\lambda_{\rm norm}=10^{-2}$,
$\lambda_{\rm init}=10^{-1}$ \\
Memory controls & Gradient checkpointing; flip chunks 16; no mixed precision \\
Reference seed & 88 with rank-dependent offsets \\
\bottomrule
\end{tabularx}
\end{table}

\subsection{Sequential optimization record}

At each segment, the optimizer resumes from the preceding model and Adam
state, draws a new deterministic minibatch stream and applies gradient
clipping after distributed gradient reduction. A complete checkpoint contains
model parameters, optimizer moments, segment index, update index, random-number
generator states and the accumulated validation history. Resuming from a
checkpoint reproduces the next minibatch and loss within the declared
floating-point tolerance.

Uniform training samples avoid a second learned sampler inside the residual
loss and make the normalization identity direct, but they can under-resolve a
highly concentrated wavefunction. The validation suite therefore evaluates
Born-weighted observables separately. Increasing the uniform batch size tests
Monte Carlo stability; it does not by itself establish accuracy in low-mass
regions. For the enumerated small cases, the exact full-basis residual is the
reference against which the minibatch estimate is checked.

The phase parameterization is continuous only modulo the network output, not a
global unwrapped phase chart. Crank--Nicolson residuals are evaluated from
complex amplitudes, so they remain insensitive to a $2\pi$ representation
change. Derivatives used for $O_\tau$ are taken before any discontinuous
post-processing. Checkpoints with saturated amplitude or phase outputs are
flagged because saturation can suppress useful derivatives even when the
pointwise amplitude error appears modest.

\subsection{Sampling and post-training queries}

At fixed $\tau$, the model distribution is
$\pi_\phi(z;\tau)=|\psi_\phi(z;\tau)|^2/\sum_{z'}|\psi_\phi(z';\tau)|^2$. A
single-spin Metropolis proposal $z\to z^{\oplus i}$ is accepted with
probability $\min\{1,|\psi_\phi(z^{\oplus i};\tau)|^2/|\psi_\phi(z;\tau)|^2\}$.
This is the symmetric-proposal Metropolis rule \cite{MetropolisEtAl1953}.
The model then supplies
\begin{equation}
E_{\rm loc}(z;\tau)=\frac{[H(\tau)\psi_\phi](z;\tau)}{\psi_\phi(z;\tau)},
\qquad O_\tau(z;\tau)=\partial_\tau\log\psi_\phi(z;\tau),
\label{eq:post-training-queries}
\end{equation}
with the derivative obtained by automatic differentiation.  The trained model
is frozen before these queries and is not retrained for candidate warps.

\subsection{MCMC retained sample and moment estimates}

The production hardness query initializes 1024 chains independently from the
uniform spin distribution, discards the first 60 single-spin proposals per
chain and retains updates $64,68,\ldots,160$. Let
$z_{cm}$ be retained state $m$ of chain $c$. For an observable $Y$, the pooled
mean and complex variance are
\begin{equation}
\widehat\mu_Y=\frac{1}{CM}\sum_{c=1}^{C}\sum_{m=1}^{M}Y(z_{cm}),\qquad
\widehat V_Y=\frac{1}{CM}\sum_{c,m}|Y(z_{cm})-\widehat\mu_Y|^2.
\label{eq:mcmc-moments}
\end{equation}
Here $C=1024$ and $M=25$. Equation~\eqref{eq:mcmc-moments} is the estimator used
for the profile, while uncertainty is estimated at the chain level to preserve
within-chain dependence.

The fixed burn-in and run length are implementation settings, not a proof of
mixing. The execution record reports acceptance rates, between-chain dispersion,
effective sample-size estimates for $E_{\rm loc}$ and $O_\tau$, and stability
under longer burn-in/run-length settings. Failed or stuck chains are not
silently removed. At small $N$, exact Born sums provide a direct bias check for
the entire MCMC-to-profile pipeline.

\section{Hardness inference and schedule construction}
\label{supp:schedule}

\subsection{Profile estimators}

The raw profile is evaluated on $G=61$ uniformly spaced $\tau$ coordinates. Exact
enumeration is used for the configured small-profile regime; larger systems
use 1024 parallel Metropolis chains, 60 burn-in proposals, thinning by four and
160 total proposals per grid point. Each chain starts from an independent
uniformly random spin configuration and is reinitialized at each point. With
updates indexed $1,\ldots,160$, states at updates $64,68,\ldots,160$ are
retained, giving 25 states per chain and 25,600 pooled states per grid point.
These settings are computational choices, not evidence of mixing; chain-count,
burn-in and thinning sensitivity are therefore included with the archived
profiles.

For retained configurations, complex moments give
\begin{align}
\widehat V_H(\tau)&=\max\{0,\E_{\pi_\phi}|E_{\rm loc}|^2
-|\E_{\pi_\phi}E_{\rm loc}|^2\},\nonumber\\
\widehat G_{\tau\tau}(\tau)&=\max\{0,\E_{\pi_\phi}|O_\tau|^2
-|\E_{\pi_\phi}O_\tau|^2\}.
\label{eq:sampled-proxies}
\end{align}
Evaluation batches contain at most 1024 configurations.  A denominator floor
of $10^{-20}$ is used only to prevent division overflow in local-energy ratios;
the number and total weight of affected samples must be recorded.  The arrays
$\{\tau_g,\Re\widehat\mu_E,\widehat V_H,\widehat G_{\tau\tau},r_\tau,
\Phi_\tau,k(\tau_g)\}$ form the raw profile record. The release also contains
the declared $F$, the pulled grid and $r(s)=F'(s)r_\tau[F(s)]$ used by the
schedule constructor.

\subsection{Equalization-only warp}

Let $R(s)=\int_0^s r(v)dv$, evaluated by cumulative trapezoidal quadrature and
monotone interpolation.  The equalization-only warp is
\begin{equation}
w_{\rm eq}(u)=R^{-1}[uR(1)],\qquad
s_j^{\rm eq}=w_{\rm eq}(j/p),\quad j=0,\ldots,p.
\label{eq:eq-warp}
\end{equation}
The identity and equalization-only schedules are retained separately.  This is
the comparison that isolates the theorem-guided allocation from subsequent
finite-depth regularization.

\subsection{Finite-depth parameterization and objective}

Candidate increments $\delta_j=s_j-s_{j-1}$ are generated from logits
$\theta_j$ by
\begin{equation}
\delta_j=\frac{f_{\min}}{p}+(1-f_{\min})
\frac{e^{\theta_j}}{\sum_{k=1}^p e^{\theta_k}},\qquad
s_j=\sum_{k=1}^j\delta_k.
\label{eq:softmax-warp}
\end{equation}
This enforces $0=s_0<\cdots<s_p=1$ and preserves the ramp endpoints in
Eq.~\eqref{eq:lr-ar-schedules}.

For a desired positive width vector $d$ with $\sum_jd_j=1$, the inverse
encoding first forms
\begin{equation}
y_j(d)=\frac{\max\{d_j-f_{\min}/p,\varepsilon_{\rm enc}\}}
{\sum_k\max\{d_k-f_{\min}/p,\varepsilon_{\rm enc}\}},\qquad
\theta_j(d)=\log y_j(d)-\frac1p\sum_k\log y_k(d),
\label{eq:inverse-logit-encoding}
\end{equation}
with $\varepsilon_{\rm enc}=10^{-12}$. Decoding these logits gives the
floor-feasible widths $f_{\min}/p+(1-f_{\min})y_j(d)$. Let
$d_j^{\rm LR}=1/p$ and $d_j^{\rm eq}=s_j^{\rm eq}-s_{j-1}^{\rm eq}$. The CEM
initial mean is
$\bm\mu^{(0)}=[\bm\theta(d^{\rm LR})+\bm\theta(d^{\rm eq})]/2$; its initial
componentwise standard deviation is the run-specific scalar given below. The
centering in Eq.~\eqref{eq:inverse-logit-encoding} fixes the softmax null
direction and is also applied after each proposal update.

Consistent with the main Article, the discrete
proxy is evaluated on the $p$ warp intervals,
\begin{equation}
\mathcal J(\bm\theta)=
\sum_{j=1}^{p}\Phi_{\rm proxy}\!\left(\frac{s_{j-1}+s_j}{2}\right)
(s_j-s_{j-1})^2+\lambda_{\rm sm}\Omega_{\rm sm}
+\lambda_{\rm LR}\Omega_{\rm LR}.
\label{eq:finite-depth-objective}
\end{equation}
The exact penalties are
\begin{align}
\Omega_{\rm sm}&=\sum_{\ell=1}^{p-1}
[(\log\gamma_{\ell+1}-\log\gamma_\ell)^2+
(\log\beta_{\ell+1}-\log\beta_\ell)^2],\nonumber\\
\Omega_{\rm LR}&=\sum_{\ell=1}^{p}[(\log\gamma_\ell-
\log\gamma_\ell^{\rm LR})^2+(\log\beta_\ell-
\log\beta_\ell^{\rm LR})^2]+
\sum_{j=1}^{p-1}(s_j-j/p)^2.
\label{eq:finite-depth-penalties}
\end{align}
All LR and AR angles in these logarithms are strictly positive under the stated
ramp convention. No post-warp angle-sum normalization is applied. The proxy is
evaluated on the $p$ layer-warp intervals, consistently with Eq.~(16) of the
main Article; cost and mixer subsegments are not independent warp variables.

The cross-entropy method \cite{Rubinstein1999} uses 50 iterations, population 96, 12 elites,
smoothing 0.7, initial logit standard deviation 0.8,
$f_{\min}=0.02$, $\lambda_{\rm sm}=0.02$ and
$\lambda_{\rm LR}=0.005$, with outer-loop seed 1234. The identity candidate is included in every
iteration.  The profile remains fixed and the final schedule is the best
regularized-proxy candidate, not the candidate with the best exact or
calibrated success probability.  Exact state-vector evaluations, where
available, are strictly post-selection diagnostics.  This separation prevents
outcome leakage into a method described as state-proxy guided.

\subsection{Cross-entropy update and decision rule}

Let $\mu_t,\sigma_t$ be the componentwise Gaussian parameters for the logits
at iteration $t$. After ranking the population by $\mathcal J$, let
$\mu_t^{\rm e}$ be the elite mean and
$\sigma_t^{\rm e}=\operatorname{Std}(\bm\theta_{\rm elite})+10^{-6}$.
The proposal update is
\begin{equation}
\mu_{t+1}=\eta\mu_t+(1-\eta)\mu_t^{\rm e},\qquad
\sigma_{t+1}=\eta\sigma_t+(1-\eta)\sigma_t^{\rm e},
\qquad \eta=0.7.
\label{eq:cem-update}
\end{equation}
The final choice is the lowest-$\mathcal J$ feasible candidate encountered
over all iterations, including the identity and equalization-only candidates.
Ties are broken deterministically by lexicographic order of the node vector.
The complete population need not be released, but the iterationwise best,
elite mean, proposal moments and random seed are required to reconstruct the
search trajectory.

The refinement is deliberately low dimensional: it has one logit per warp
interval and does not optimize independent $\gamma$ and $\beta$ angles. The
shared warp preserves the connection to the learned path. Because the softmax
is invariant under adding a constant to every logit, the implementation
centres $\bm\theta$ after each update. This removes a null direction without
changing any candidate schedule.

The LR penalty is zero for the identity and does not
enforce equality of total angle sums. Hyperparameters are fixed before hardware
execution and are not re-tuned separately for LR and AR.

Reported outcome tables use schedules generated with the $p$-interval,
no-angle-renormalization, proxy-only selection and CEM conventions defined
above. The archived nodes, angles, CEM history and bound circuits record these
choices.

\subsection{Discrete consistency checks}

Every candidate is rejected unless its nodes are finite, strictly increasing,
within $[0,1]$ and end exactly at one within machine precision. Substituting
the nodes into Eq.~\eqref{eq:lr-ar-schedules} must reproduce the declared
endpoints. The compiler input is hashed after angle serialization to detect
unit or ordering changes between schedule construction and execution.

Quadrature convergence is assessed by recomputing $R$, $w_{\rm eq}$ and the
selected proxy candidate on denser grids. The diagnostic is the maximum node
shift, not merely the change in the integrated proxy, because a localized
interpolation error can move a layer while leaving the integral nearly
unchanged. Boundary-adjacent grid points use the one-sided convention of
\ref{supp:theory}. No smoothing operation is allowed to mix
the cost-side and mixer-side value at a discontinuity.

\subsection{Frozen execution record}

The execution record for every reported pair contains the LR,
equalization-only and refined nodes; their angle vectors; the proxy and penalty
components; the CEM history and seed; and the time at which the schedule was
frozen. No quantum
hardware counts, hardware gradients or repeated hardware objective evaluations
enter profile estimation or candidate selection.

\section{Verification and proxy validation}
\label{supp:validation}

This note distinguishes deterministic checks, quantitative validation records
and end-to-end outcomes. A diagnostic is not described as ``validated'' unless
its numerical value is included in the source data.

\subsection{Deterministic implementation checks}

The following checks are exact up to floating-point rounding: (i) Eq.
\eqref{eq:ising-map} reproduces $C(b)$ on enumerated small systems and sampled
large-system strings; (ii) the local action in Eq.~\eqref{eq:local-h-action}
matches a dense-matrix action at small $N$; (iii) different chunk sizes give
the same moments; and (iv) direct exponentiation of the piecewise Hamiltonian
matches the cost-then-mixer circuit convention. Their relative residuals and
test seeds are retained together with the pass/fail flags.

\subsection{State and dynamics checks}

For exactly propagated reference states, normalize the twin state and compute
\begin{align}
F_\psi(\tau)&=|\braket{\psi_{\rm ex}(\tau)|\widetilde\psi_\phi(\tau)}|^2,\nonumber\\
D_{\rm TV}(\tau)&=\frac12\sum_z
\left||\braket{z|\widetilde\psi_\phi(\tau)}|^2
-|\braket{z|\psi_{\rm ex}(\tau)}|^2\right|.
\label{eq:state-validation}
\end{align}
Training constraints are checked on fresh samples using a normalized
Crank--Nicolson residual, the rescaled normalization error and the initial-state
error.  At segment-interior points, a differential residual
$D_\phi=\partial_\tau\psi_\phi+\ii T H_k\psi_\phi$ directly tests the learned
coordinate dependence. Validation coordinates cover every segment,
because the sequential training scheme can otherwise hide forgetting.

The normalized Crank--Nicolson residual at segment $k$ is
\begin{equation}
\epsilon_{{\rm CN},k}=
\frac{\left\|(I+\ii\Delta t_kH_k/2)\widetilde\psi_\phi(\tau_{k+1})-
(I-\ii\Delta t_kH_k/2)\widetilde\psi_\phi(\tau_k)\right\|_2}
{\|\widetilde\psi_\phi(\tau_k)\|_2+\|\widetilde\psi_\phi(\tau_{k+1})\|_2},
\label{eq:cn-validation}
\end{equation}
using full vectors where enumeration is feasible and an independently sampled
estimate otherwise. Observable errors are reported as
$|\langle H\rangle_\phi-\langle H\rangle_{\rm ex}|$ and
$|V_{H,\phi}-V_{H,{\rm ex}}|$ in both absolute and scale-normalized units.
Fidelity alone is insufficient because a small componentwise error can affect
a derivative or a small variance disproportionately.

At segment boundaries, state fidelity and TV distance remain well defined but
the differential residual and score require one-sided evaluation. The report
therefore labels every validation coordinate by its segment and side. A
coordinate exactly on a boundary is never silently averaged across the two
Hamiltonians.

\subsection{Profile and schedule checks}

The exact and learned profiles are compared after unit-mass normalization,
$\bar r(s)=r(s)/\int_0^1r(v)dv$.  Appropriate records include an $L^1$ profile
error, Pearson and Spearman correlations, the maximum cumulative-coordinate
error and the root-mean-square node displacement,
\begin{equation}
\epsilon_w=\left[\frac{1}{p-1}\sum_{j=1}^{p-1}
(s_{j,\phi}-s_{j,{\rm ex}})^2\right]^{1/2}.
\label{eq:warp-error}
\end{equation}
Metropolis estimates are compared with enumeration at small $N$, and the
identity $\widehat G_{\tau\tau}\simeq T^2\widehat V_H$ is checked away from segment
boundaries.  Grid sizes and sampling seeds are varied until the node
displacement is small relative to the typical interval $1/p$.

For normalized profiles $\bar r_1,\bar r_2$, the reported distances are
\begin{equation}
d_1=\int_0^1|\bar r_1(s)-\bar r_2(s)|\,ds,\quad
d_R=\max_s\left|\int_0^s[\bar r_1(v)-\bar r_2(v)]\,dv\right|,
\quad d_\infty=\max_s|\bar r_1(s)-\bar r_2(s)|.
\label{eq:profile-distances}
\end{equation}
The cumulative distance $d_R$ is especially relevant because it directly
controls the inverse-CDF construction. Rank correlation is reported only when
the profile is not numerically constant. Confidence intervals for Monte Carlo
profile quantities are formed by resampling chains, not individual retained
states.

The outcome comparison is performed only after all profile and schedule files
are frozen. Proxy/outcome rank correlation may then be reported descriptively
over the already generated CEM population if all candidates are evaluated by
the same independent simulator budget, but it cannot be used to replace the
predeclared selection rule. This distinction prevents an apparently
``validation-only'' analysis from becoming hidden outcome-based tuning.

\subsection{Validation criteria and reporting}

A validation claim requires numerical diagnostic values and prespecified
tolerances. Tolerances are defined relative to the downstream node spacing and
objective uncertainty, and each reported instance is accompanied by the
complete validation record and any failed criterion. A failed diagnostic need not invalidate every
hardware observation, but it limits claims that the observation was caused by
an accurately learned proxy.

Validation data are partitioned by purpose. Values used to choose architecture,
loss weights, $a,b,\epsilon$, penalties or CEM settings belong to a calibration
set; values used only after freezing belong to a held-out audit set. Reporting
small-$N$ exact outcomes during tuning and later calling them validation would
introduce post-selection. The release manifest therefore timestamps the
configuration, profile, schedule and outcome stages.

\section{Large-scale implementation}
\label{supp:hpc}

\subsection{Distributed training and local workload}

\TxNQDT{} training and proxy extraction are performed on distributed NVIDIA
A100 GPU resources, with one model replica per process and data-parallel
gradient averaging. Each rank samples an
independent but reproducible minibatch; gradients are synchronized before the
Adam update.  The sparse diagonal term in Eq.~\eqref{eq:local-h-action} scales
with the retained interaction list, while the transverse term requires $N$
neighbour amplitudes per sampled configuration.  Batching, flip chunking and
gradient checkpointing make memory polynomial in the sampled batch, $N$ and
model width.  This is not a polynomial-time guarantee for learning arbitrary
quantum states, nor is the large-$N$ workflow an exact state-vector simulator.

Final schedule construction, circuit generation and postprocessing are
performed on an NVIDIA GH200 Grace Hopper system. Once the raw arrays on
$\tau_g$ and their pulled ramp-grid profile $r(s_g)$ have been stored,
candidate evaluation no longer requires transformer inference. Software,
hardware and run-specific configurations are retained with the computational
artifacts. The final output is a fixed logical circuit specification; the
quantum processor is not part of the training loop.

\subsection{Parallelization and memory accounting}

The distributed job uses data parallelism rather than model parallelism. Each
rank holds a complete network and evaluates a disjoint portion of the global
minibatch; an all-reduce forms the global gradient. Consequently, adding ranks
reduces per-rank sample memory but replicates parameters and optimizer states.
Reported memory therefore includes model, Adam moments, activations,
single-flip neighbour chunks and communication buffers. Peak allocated and
peak reserved accelerator memory are distinguished.

For a local batch $B_r$ and flip chunk $c$, the dominant inference tensor
contains $O(B_r c)$ configurations at a time, repeated over $N/c$ chunks. This
explains how the local Hamiltonian can be evaluated without a $B_rN$-sized
activation graph, but it does not make the total arithmetic independent of
$N$. The diagonal interaction sum is evaluated from the exact sparse edge list.
The release records the edge count because it affects runtime independently of
the number of variables.

Deterministic reproduction across different world sizes is not assumed:
floating-point reductions and minibatch partitioning can change optimization
trajectories. Reproducibility means that a declared environment, world size and
seed set reconstructs the archived run, while sensitivity runs quantify the
effect of changing those choices.

\subsection{Numerical safeguards and provenance}

Every training run records segment losses, gradient norms, elapsed times and
finite-value checks for amplitudes, Hamiltonian actions, residuals and
gradients.  The CPU checkpoint contains model and optimizer states,
architecture and QAOA configurations, segment boundaries, random-generator
states and environment metadata.  The run-level chain is
\begin{equation}
Q\longmapsto \text{checkpoint}\longmapsto r_\tau(\tau_g)
\longmapsto r(s_g)\longmapsto \{s_j\}
\longmapsto (\bm\gamma^{\rm AR},\bm\beta^{\rm AR}),
\label{eq:provenance-chain}
\end{equation}
and each arrow must be recoverable from saved machine-readable data.

Hashes are computed for the QUBO, segment table, checkpoint, profile, node and
angle files. A manifest links those hashes to software version, container or
environment lock file, accelerator type, compiler flags and precision policy.
The serialized checkpoint is treated as an intermediate scientific artifact,
not as sufficient documentation by itself: source code and architecture
configuration are required to interpret it.

Crashes, non-finite losses and restarts are retained in the run log. If a run
is restarted from an earlier checkpoint, the manifest records the discarded
updates and the new seed state. Manual selection among multiple training runs
based on downstream QAOA performance would constitute outcome-guided model
selection and must either be prohibited or disclosed with a held-out test.

\section{IBM quantum-hardware implementation and fairness audit}
\label{supp:hardware}

The hardware protocol is the one stated in the main Article:
\texttt{ibm\_pittsburgh}, Qiskit Runtime \texttt{SamplerV2}, transpiler
optimization level 3, the common line-topology prescription with
$k=2$ (denoted $k_{\rm cut}$ below), XpXm dynamical decoupling and common
gate/measurement Pauli twirling
\cite{JavadiAbhariEtAl2024,WallmanEmerson2016,EzzellEtAl2023}. LR and AR use
the same measurement budget and are submitted within the same batch whenever
possible. The adaptive schedule is fixed before hardware submission.
Backend job metadata, rather than prose alone, are the authoritative record of
these settings.

\subsection{Logical circuit and interaction support}

Logical terms are implemented as
\begin{align}
e^{-\ii\gamma h_iZ_i}&=R_Z(2\gamma h_i), &
e^{-\ii\gamma J_{ij}Z_iZ_j}&=R_{ZZ}(2\gamma J_{ij}), &
e^{+\ii\beta X_i}&=R_X(-2\beta).
\label{eq:logical-gates}
\end{align}
The LR and AR circuits use the same ordered logical variables, fields,
couplings and cost--mixer block template. The logical variables are placed in
the declared line order. The logical cost Hamiltonian is constructed from the complete archived QUBO
defined in Supplementary Note~2. Every nonzero off-diagonal coefficient is
retained in Tx-NQDT training, reference-minimum determination, circuit
construction and objective evaluation. The parameter $k_{\mathrm{cut}}=2$
belongs to the declared hardware-routing and compilation prescription; it does
not truncate the logical QUBO. The Transformer graph mask is a separate
architectural choice: direct token-to-token attention follows the one-hop QUBO
graph, while the local-energy estimator retains the complete interaction list. The resulting
implemented QUBO defines all decoded objective values, minima and accepted
sets, and its exact ordered interaction list is archived.

The sign and factor conventions in Eq.~\eqref{eq:logical-gates} are verified
on one- and two-qubit matrices before transpilation. A logical term table
contains the ordered qubit indices, coefficient, bound angle and source QUBO
entry. The table is shared by LR and AR; only the scalar schedule factors
change. Terms with numerically zero angles are retained in the logical audit
even if a compiler removes the corresponding gate.

Distance-two $ZZ$ terms can still require routing or backend-native synthesis.
The same ordered term table is used for LR and AR and is hashed at the training,
schedule, circuit and decoding stages.

\subsection{Compilation and execution}

Members of a matched pair use the same backend target,  the IBM Quantum
Heron R3 \texttt{ibm\_pittsburgh} processor accessed through IBM Quantum Compute
\texttt{SamplerV2}, routing prescription,
transpiler configuration in Qiskit's transpiler and requested shots, and are submitted within the
same batch whenever possible.  The adaptive angles are frozen before circuit
generation. For each compiled circuit, the archive contains the physical layout, logical and
physical depths, one- and two-qubit gate counts, measurement map and compiled
duration when available.  Equal logical depth alone does not guarantee equal
physical duration or noise exposure, so these values are reported rather than
assumed equal.

Transpilation is performed before the execution comparison is inspected. If a
randomized transpiler seed is used, the same prespecified seed set is applied
to both schedules. Choosing the shallowest compiled AR circuit and an arbitrary
LR circuit would bias the comparison. The primary pair either uses the same
seed or a prespecified aggregate across paired seeds; both bound circuits and
their transpiled instruction streams are archived.

Angle-dependent optimizations can cause different gate cancellations even
under a common template. The fairness audit therefore reports both logical and
physical resources and does not force them to be equal after compilation. A
resource-adjusted analysis may be informative, but the headline comparison
remains the prespecified compiler output because manually padding one circuit
would introduce a different noise process.

XpXm dynamical decoupling and the same Pauli-twirling policy are applied to
both schedules. No zero-noise extrapolation or output-level quasiprobability
reconstruction was applied. Any readout-corrected or
classically improved result is labelled separately and does
not replace the raw-count comparison \cite{NationEtAl2021}.

When randomized or interleaved execution is enabled, its order is archived.
Available calibration snapshots before and after a batch include the
relevant qubit readout errors, one- and two-qubit gate errors, coherence times
and backend timestamp. These records do not fully characterize nonstationary
noise, and a single calibration window remains one experimental replicate.

Requested and returned shots are read from job metadata and counts,
respectively; any
discrepancy is reported. Sampler options such as twirling shots per randomization
and number of randomizations are also retained because they determine how the
total budget is distributed.

Ongoing work extends this protocol with untruncated cost operators,
deterministic SAT-optimized routing, matched-pair compilation verification,
control schedules and error suppression routines. Results from this extended campaign will be reported in an updated version of this manuscript.

\subsection{Decoding and raw probabilities}

Archived classical strings are mapped to logical variables using the saved
classical-bit and physical-layout maps before evaluating Eq.~\eqref{eq:qubo}.
For counts $n_b$,
\begin{equation}
S_{\rm eff}=\sum_b n_b,\qquad
\widehat p_{\rm hw}(b)=\frac{n_b}{S_{\rm eff}},\qquad
\widehat P_{\rm succ}(\rho)=\frac{1}{S_{\rm eff}}
\sum_{b\in A_\rho}n_b.
\label{eq:hardware-probabilities}
\end{equation}
The returned denominator is used even if it differs from the request.  Since
probabilities near $1/S_{\rm eff}$ can have very large relative gains, the
source table includes threshold counts and effective shots, not only
percentages.

Bit-order validation uses basis-state test circuits or an independently
constructed decoder test. Qiskit display order, classical-register order,
physical qubit order and logical variable order are not assumed identical.
The required decoder check evaluates strings with known objective values; its
test output and a hash of the mapping accompany every count file.

Raw expected cost and variance are computed directly from counts,
$\widehat{\E[C]}=S_{\rm eff}^{-1}\sum_b n_b C(b)$ and
$\widehat{\Var(C)}=S_{\rm eff}^{-1}\sum_b n_b[C(b)-\widehat{\E[C]}]^2$.
The best observed objective is the minimum over nonzero counts. Threshold
counts are then accumulated from the same decoded table, ensuring that mean,
best and success probabilities cannot be based on different post-processing
or interaction conventions.

\subsection{Scope of the hardware inference}

The hardware experiment estimates schedule-dependent output frequencies on the
reported device and calibration context.  It does not establish device independence or temporal reproducibility across calibrations.  A stronger
hardware claim would require independent repeats across randomized execution
order, compilation seeds and calibration windows.  Shot resampling of one
archived job quantifies measurement noise but cannot substitute for those
independent repeats.

There are three distinct uncertainty layers: multinomial shot noise within a
job, randomized compilation/twirling variation within a calibration window,
and device drift across windows. The Clopper--Pearson intervals in
\ref{supp:statistics} address only the first layer for a
fixed circuit and accepted set. Where multiple randomized circuits are
available, results are first reported per circuit and then aggregated using a
prespecified hierarchical or cluster-aware procedure. Treating all shots from
different circuits as independent Bernoulli trials would understate
between-circuit variation.

\section{Benchmark design and statistical analysis}
\label{supp:statistics}

\subsection{Evidence classes and reference provenance}

Every result is labelled as exact state-vector, hardware-observed or, if used
only for exploration, model-predicted.  An enumerated or solver-certified
minimum is distinguished from a best-known retained reference.  The latter
defines a reproducible threshold set but does not justify the phrase ``exact
optimum''.  

The principal design is paired within each $(N,p)$ configuration but includes
only the reported benchmark instances.  Consistency across sizes and depths is
useful evidence; it is not a population-level statement over the QUBO
distribution.  Claims of generic or scalable superiority require multiple
independent instances at each size and a prespecified aggregation rule.

\subsection{Effect measures}

For a matched pair, report
\begin{align}
\Delta P(\rho)&=P_{\rm succ}^{\rm AR}(\rho)-P_{\rm succ}^{\rm LR}(\rho),\nonumber\\
\Delta_{\rm rel}(\rho)&=\frac{P_{\rm succ}^{\rm AR}(\rho)
-P_{\rm succ}^{\rm LR}(\rho)}{P_{\rm succ}^{\rm LR}(\rho)},
\qquad F_{\rm AR/LR}=1+\Delta_{\rm rel}.
\label{eq:effect-measures}
\end{align}
Relative changes are undefined when the LR estimate is zero and are unstable
near the resolution floor.  The absolute probabilities, counts and
$\Delta P$ are therefore primary; fold and percentage changes are secondary.
The lowest observed objective
$C_{\rm best}^{\rm obs}=\min_{b:n_b>0}C(b)$ is reported with $S_{\rm eff}$ and
is interpreted as an extreme statistic, not an estimate of expected quality.

For a fixed distribution with cumulative objective distribution
$F_C(c)=\Pr[C(b)\le c]$, the minimum of $S$ independent draws obeys
$\Pr[C_{\min}>c]=[1-F_C(c)]^S$. This relation makes explicit why best-observed
values cannot be compared at unequal shot counts without adjustment. When the
full exact distribution is available, the expected minimum at the hardware
shot budget can be computed as a secondary simulator quantity; on hardware,
the raw order statistic and shot count remain the transparent primary record.

The expected objective can be converted to an approximation ratio only under
a declared normalization that handles the worst/reference objective and is
invariant to irrelevant shifts. Because the Article reports negative QUBO
energies and threshold percentages relative to $C^\star$, the analysis keeps
the raw expected energy and accepted-set probability rather than inventing a
new ratio after observing the data.

\subsection{Finite-shot intervals and nested thresholds}

Conditional on a fixed compiled circuit, the accepted count obeys
$K_\rho\sim\mathrm{Binomial}(S_{\rm eff},P_{\rm succ}(\rho))$.  Two-sided 95\%
Clopper--Pearson intervals are used, including their exact boundary forms
\cite{ClopperPearson1934}.  For $0<K<S$,
\begin{equation}
[P_{\rm lo},P_{\rm hi}]=
\left[B^{-1}\!\left(0.025;K,S-K+1\right),
B^{-1}\!\left(0.975;K+1,S-K\right)\right].
\label{eq:clopper-pearson}
\end{equation}
For the LR--AR difference, use a conservative exact Bonferroni construction.
Let $[L_A,U_A]$ and $[L_L,U_L]$ be two-sided 97.5\% Clopper--Pearson
intervals for AR and LR, respectively. The reported 95\% interval is
\begin{equation}
[\Delta_L,\Delta_U]=[L_A-U_L,\;U_A-L_L].
\label{eq:exact-difference-interval}
\end{equation}
Its simultaneous coverage is at least 95\% by the union bound. It is
conservative but remains valid at zero counts and unequal returned-shot totals;
overlap of two ordinary 95\% marginal intervals is not used as a test.

LR and AR shot outcomes from separate circuit executions are not paired at the
shot level even when the jobs share a batch. Conditional on the two fixed
circuits, an independent-binomial variance estimate for the absolute
difference is
\begin{equation}
\widehat{\operatorname{Var}}(\widehat\Delta P)=
\frac{\widehat P_{\rm AR}(1-\widehat P_{\rm AR})}{S_{\rm AR}}+
\frac{\widehat P_{\rm LR}(1-\widehat P_{\rm LR})}{S_{\rm LR}},
\label{eq:difference-variance}
\end{equation}
but a normal interval is unreliable at very small counts. Therefore
Eq.~\eqref{eq:exact-difference-interval} is used consistently across
thresholds. Zero-count
outcomes are retained and reported with one-sided resolution bounds rather
than replaced by pseudocounts in the headline table.

The accepted sets are nested, so the five threshold estimates are correlated.
If $A_{\rho_a}\subseteq A_{\rho_b}$, then
\begin{equation}
\operatorname{Cov}\!\left[\widehat P(\rho_a),\widehat P(\rho_b)\right]
=\frac{P(\rho_a)[1-P(\rho_b)]}{S_{\rm eff}}.
\label{eq:nested-covariance}
\end{equation}
The threshold profile is therefore one structured outcome, not five
independent confirmations.  Negative changes at individual thresholds and
worse best-observed energies are retained; ``consistent improvement'' is not
used to hide such cases.

If a simultaneous claim across thresholds is made, multiplicity and nesting
must be handled jointly, for example by inverting a multinomial confidence
region over the disjoint objective bands. The current tables are principally
descriptive and report all predeclared thresholds, so no threshold is selected
post hoc as the sole success criterion. Rounding is applied only after counts,
probabilities, intervals and differences are calculated at full precision.

\subsection{Reporting hierarchy and missingness}

The primary hardware table lists, for every LR--AR pair and threshold, returned
shots, accepted counts, absolute probabilities, $\Delta P$ and its interval.
Relative improvement is displayed only when the baseline is nonzero and above
the declared resolution floor. Mean objective and best observed value are
reported once per circuit. This ordering prevents a large relative percentage
based on one or two counts from dominating a more informative absolute result.

Jobs or circuits are excluded only by prespecified technical criteria such as
backend failure or an undecodable count register, never because of an
unfavourable outcome. All exclusions, retries and partial returns remain in a
flow table with job identifiers. A retried pair is rerun symmetrically when
possible; combining a favourable member of one attempt with the other member
of a later attempt is prohibited.

\subsection{No outcome-based schedule selection}

The reported schedule is selected solely by the frozen regularized proxy in
Eq.~\eqref{eq:finite-depth-objective}; exact simulations and hardware counts
are evaluated afterward. An outcome-guided variant requires a separate
validation/holdout design and is labelled as a different algorithm.


\section{Reproducibility and released artifacts}
\label{supp:reproducibility}

\subsection{Release contents}

The machine-readable archive contains the files in Table~\ref{tab:release} and
is identified by the persistent repository identifier reported in the Data
availability statement.

\begin{table}[htbp]
\centering
\caption{Minimum machine-readable release.}
\label{tab:release}
\small
\begin{tabularx}{\textwidth}{L{0.28\textwidth}Y}
\toprule
Group & Contents \\
\midrule
Instances & QUBO matrices, seeds, Ising maps, implemented hardware term lists,
reference values and certification records \\
Training & Source code, model configuration, checkpoints, loss/residual
histories, RNG states and environment record \\
Profiles & Grid, sample settings, energy/score moments, hardness arrays,
segment indices and validation diagnostics \\
Schedules & LR, equalization and refined nodes/angles, CEM history, objective
components and freeze metadata \\
Exact calculations & State-vector code, exact probabilities and candidate
diagnostics that were not used for selection \\
Hardware & Logical/transpiled circuits, layouts, options, job metadata, raw
counts and decoding maps \\
Source data & Values underlying every main-text and supplementary table,
including counts and intervals \\
Environment & Package lock file, accelerator/runtime versions and SHA-256
manifest \\
\bottomrule
\end{tabularx}
\end{table}

\subsection{Directory structure and immutable identifiers}

The archive uses the following directory structure:
\begin{verbatim}
instances/       q matrices, ising maps, reference certificates
training/        source, configs, checkpoints, logs, rng states
profiles/        grids, samples summaries, moments, validation
schedules/       lr/equalized/refined nodes, angles, cem histories
exact/           exact distributions and post-freeze diagnostics
hardware/        circuits, layouts, job metadata, counts, decoders
analysis/        table builders, intervals, source-data exports
manifests/       checksums, environment locks, licenses, citation
\end{verbatim}
Each reported configuration receives a stable identifier derived from the
instance checksum, $p$, model configuration, profile configuration and schedule
search configuration. Hardware executions add the backend, compilation seed
and job identifier. Human-readable filenames may change, but the manifest
retains these immutable links.

Raw counts are never overwritten by corrected or aggregated tables. Derived
files contain parent hashes and the analysis commit. The same rule applies to
profiles: smoothing, normalization and perturbation variants are separate files
with explicit parents. This provenance prevents a later analysis from silently
changing the schedule input while preserving the same filename.

\subsection{Reconstruction checks}

The preferred reproduction path loads the archived checkpoint; stochastic
retraining is a separate, stronger test.  A successful reconstruction verifies
the QUBO--Ising identity, segment convention, profile arrays, monotone nodes,
angle endpoints and count totals.  In particular,
\begin{equation}
C(b)=c_0+E_z(z),\qquad s_0=0<s_1<\cdots<s_p=1,
\qquad S_{\rm eff}=\sum_b n_b.
\label{eq:reconstruction-identities}
\end{equation}
The reconstructed threshold counts and probabilities are compared with the
source tables.  Bitwise equality is not required for stochastic retraining;
state, profile and warp diagnostics in Supplementary Note~6 define the relevant
tolerances.

Two reproduction levels are distinguished. Artifact reconstruction starts
from archived checkpoints and counts and reproduces the published
profiles, schedules and tables. Independent rerun reproduction begins from
source and random seeds and retrains/re-executes; it can differ statistically
because of optimization and hardware variation. Both are valuable, but only
the first can be required to match the exact archived numerical values.

Artifact integrity is verified by automated checks for QUBO checksum
mismatches, nonmonotone node vectors, endpoint mismatches, citation-key errors,
count-total mismatches and table values not traceable to source data. The checks
also flag profile nonconvergence, poor MCMC diagnostics, large
physical-resource imbalance and missing device metadata. Archived build
metadata record the source commit and bibliography checksum.

\subsection{Availability statement}

Code, trained checkpoints, instances, schedules, circuit metadata and raw
counts will be deposited in a persistent public repository with the published
work. The archival record identifies the DOI, license, code commit, data
version and access date. Hardware
counts reproduce the archived analysis but do not imply that a future device
execution will reproduce the same calibration-dependent distribution.

\section{Scope, limitations and generalization}
\label{supp:scope}

\subsection{What is established}

The mathematical result is the optimizer of
Eq.~\eqref{eq:proxy-functional} on a fixed one-dimensional path.  The empirical
result is a set of matched LR--AR comparisons for the reported sparse-QUBO
instances, depths and hardware context.  The study does not establish a
complexity advantage over classical optimization, universal AR superiority,
device independence, reduced physical runtime, spectral-gap recovery or
population-level performance over random QUBOs.

The \TxNQDT{} is an approximate classical surrogate trained separately for a
reference trajectory.  Avoiding explicit $2^N$ storage does not remove the
possibility of difficult optimization, poor MCMC mixing or representation
error.  Large-$N$ claims therefore rely on the retained residual, sampling,
profile and schedule diagnostics, together with frozen-circuit hardware
evaluation.

\subsection{General positive proxies}

The theorem accepts any positive integrable density.  A general construction
may use
\begin{equation}
r(s)=\epsilon+\sum_{m=1}^{M}c_m\sqrt{\max\{P_m(s),0\}},
\qquad c_m\ge0,
\label{eq:general-proxy}
\end{equation}
where $P_m$ is a state-, model- or hardware-derived diagnostic.  Components
with different units require explicit standardization and coefficients fixed
on independent development data.  Hardware-aware terms change the scientific
question from state-informed logical allocation to joint logical--physical
control and should be evaluated as a distinct method.

\subsection{Beyond the present QAOA path}

For a fixed curve $\bm\lambda(u)$ in a multidimensional control space, an
induced positive line element can be equalized in the same manner.  Optimizing
the curve itself is a different problem connected to geometric optimal control
and is not covered by Supplementary Theorem~1.  Shared warps may also be
applied to other alternating-operator circuits when their reference parameter
families and local state estimators are well defined.  Such extensions require
new validation; they do not follow empirically from the QUBO benchmarks.

The present workflow is offline: the twin is trained, the profile inferred and
the schedule frozen before device execution.  Online updates using hardware
outcomes would consume an additional quantum evaluation budget and must be
compared under a correspondingly expanded resource accounting.

\bibliographystyle{unsrtnat}
\bibliography{supplementary}